\documentclass[11pt,a4paper]{article}

\usepackage[utf8]{inputenc}
\usepackage[T1]{fontenc}

\usepackage[authoryear]{natbib}
\usepackage{hyperref}

\usepackage{amsmath}
\usepackage{amsfonts}
\usepackage{amssymb}

\usepackage{graphicx}

\usepackage[ruled,vlined]{algorithm2e}

\SetInd{0.15cm}{0.25cm}
\DecMargin{0.8\parindent}
\DontPrintSemicolon
\SetKwFor{AtPos}{At time}{:}{end}
\SetKw{Sample}{Sample}
\SetKw{Compute}{Compute}
\SetKw{Resample}{Resample}
\SetKw{Approximate}{Approximate}
\SetKw{From}{from}
\SetKw{And}{and}
\SetKw{To}{to}
\SetKw{Set}{Set}
\SetKw{Denote}{Denote}
\SetKw{KwWith}{with}
\SetKwFor{For}{For}{:}{end}
\SetKwFor{With}{With}{:}{end}
\SetKwIF{If}{ElseIf}{Else}{if}{:}{else with}{else}{end}



\newcommand{\R}{\mathbb R}
\newcommand{\N}{\mathbb N}

\newcommand{\expect}[1][]{\mathbb E_{#1}}
\newcommand{\var}[1][]{\mathbb V_{#1}}

\newcommand{\chitwo}[1][]{\mathcal D\chi^2_{#1}}

\newcommand{\Algo}{\mathcal A}
\newcommand{\M}{\mathbb M}
\newcommand{\U}{\mathbb U}
\newcommand{\K}{\mathbb K}
\newcommand{\spaceX}{\mathcal X}
\newcommand{\spaceY}{\mathcal Y}

\newcommand{\tmin}{t_{\mathrm{start}}}
\newcommand{\tmax}{t_{\mathrm{end}}}

\newcommand{\refer}[1]{#1^{\mathrm{ref}}}
\newcommand{\ancestor}[1]{\boldsymbol{\tilde{#1}}}

\newcommand{\Ntrain}{N_{\mathrm{train}}}
\newcommand{\Nsample}{N_{\mathrm{sample}}}

\newcommand{\one}{\mathbf{1}}

\DeclareMathSymbol{:}{\mathord}{operators}{"3A}

\begin{document}
\title{A forward-only scheme for online learning of proposal distributions in particle filters}
\date{}

\author{
  Sylvain Procope-Mamert\thanks{Universit\'{e} Paris-Saclay, INRAE, MaIAGE, 78350, Jouy-en-Josas, France}
  \and Nicolas Chopin\thanks{ENSAE, CREST, Institut Polytechnique de Paris}
  \and Maud Delattre\footnotemark[1]
  \and Guillaume Kon Kam King\footnotemark[1]
}

\maketitle

\begin{abstract}
  We introduce a new online approach for constructing proposal distributions in particle filters
  using a forward scheme. Our method progressively incorporates future observations to refine proposals.
  This is in contrast to backward-scheme algorithms that require access to the entire dataset,
  such as the iterated auxiliary particle filters \citep{guarniero_iterated_2017}
  and controlled sequential Monte Carlo \citep{heng_controlled_2020},
  which leverage all future observations through backward recursion.
  In comparison, our forward scheme achieves a gradual improvement of proposals
  that converges toward the proposal targeted by these backward methods.
  We show that backward approaches can be numerically unstable even in simple settings.
  Our forward method, however, offers significantly greater robustness with only a minor trade-off
  in performance, measured by the variance of the marginal likelihood estimator.
  Numerical experiments on both simulated and real data illustrate the enhanced stability of our forward approach.
\end{abstract}

\section{Introduction}

Sequential Monte Carlo (SMC) methods \citep{doucet_tutorial_2011, chopin_introduction_2020} are a class of particle algorithms
designed to approximate sequences of probability measures known up to a normalizing constant
with a sequence of weighted particles.
They are widely used for inference in \emph{state-space models} (SSMs),
a class of models describing the evolution of an unobserved latent process through time, combined with a noisy observation process \citep{durbin_time_2012}.
An SSM typically consists of a latent Markov process and an observation process conditionally independent given the states.
These models are particularly useful when observations are highly noisy,
but the underlying dynamics provide essential structure to link observations together
and share information across time, thereby reducing uncertainty and improving estimation.
SSMs are ubiquitous in statistics, with applications in econometrics, signal processing, epidemiology, and robotics
\citep{cappe_inference_2005}.
However, inference in nonlinear and non-Gaussian SSMs is challenging because the likelihood,
the filtering, and the smoothing distributions lack closed-form expressions and are high-dimensional.
While exact solutions exist for linear Gaussian cases via the Kalman filter \citep{kalman_new_1960},
Monte Carlo methods, such as particle filters and sequential Monte Carlo methods
were introduced to handle more general settings.

SMC algorithms may be naturally derived for most SSMs from the sequential nature of the model.
The simplest and canonical SMC algorithm used to infer an SSM is the bootstrap particle filter \citep{gordon_novel_1993}.
This algorithm is often subject to particle degeneracy,
resulting in a poor approximation of the target distribution when one or a few particles carry most of the weight.
Designing efficient and robust filtering algorithms is crucial,
as poor approximations and particle degeneracy lead to high variances for Monte Carlo estimates
and unreliable inference, particularly in complex models.
There have been many attempts to develop improved SMC algorithms
that adapt to the target distribution, i.e.,\ to the model and the observed data.
For example, some authors have proposed guided particle filters \citep{doucet_sequential_2000}
or auxiliary particle filters \citep{pitt_filtering_1999}.
Most of these methods involve looking one step (one observation) ahead to sample and weight particles at each time step.
There are various ways to construct such SMC algorithms manually \citep{chopin_introduction_2020},
but they require a deep understanding of the target distributions (for example, using Taylor expansions).
Automatic and adaptive particle filters aim to reduce manual tuning and improve reliability across diverse applications,
making them an active area of research.

One approach to automatically construct better proposals is to iterate forward
and backward passes over the data in order to progressively refine the
proposals, as the iterated auxiliary particle filter (IAPF) of
\citet{guarniero_iterated_2017}, and  the controlled sequential Monte Carlo
(cSMC) of  \citet{heng_controlled_2020}. We observe that, in moderately
challenging cases, these methods may fail to produce low-variance estimates of
the marginal likelihood of the SSM. One explanation is that it is often
difficult to design the initial forward pass that provides a good
initialization to the next backward pass. This happens, for instance, if one
uses the bootstrap filter for this initial forward pass.

In this article, we propose a forward scheme
that produces an SMC algorithm without requiring a good initialization.
The difference with backward schemes can be understood as follows:
the backward schemes start from the last time step of the SSM and go backward in time, essentially using all future observations to construct the proposals,
whereas our forward scheme starts one step ahead and proceeds forward in time, gradually incorporating future observations to refine the proposals.
Thus, our forward scheme iterates over easier steps than the backward schemes and is much more robust to difficult initializations.
Although our scheme can be less efficient in easy cases
— it often converges more slowly than backward schemes that incorporate all available information —
we find this loss is modest.
In Markovian models, observations far ahead typically carry little information about the current state,
which limits the performance gap.
Through extensive illustrations on simulated and real data, we observe that
our algorithm is able to produce lower variance estimates than backward algorithms,
which suffer from a lack of robustness.

A recent proposal \citep{xue2025onlinerollingcontrolledsequential} uses the same forward decomposition
to construct an online smoothing algorithm, relying on two simultaneous particle systems.
The implementation and goals differ from ours:
they focus on estimating smoothing distributions rather than the marginal likelihood and parameter inference.
Nonetheless, this work demonstrates that the forward decomposition can be used fruitfully to construct adaptive particle algorithms for SSMs.

Building on these ideas, we now present our contribution in detail.
The remainder of the article is organized as follows.
In the next section, we introduce notation, state-space models, and the Feynman-Kac formalism underlying SMC algorithms.
Next, we introduce auxiliary Feynman-Kac models and characterize globally and locally optimal proposals.
Then, we derive our forward-only iterative scheme and discuss its practical implementation.
The last section reports numerical experiments on simulated and real data,
comparing our method to existing backward-scheme algorithms and to a reference bootstrap particle filter.
We conclude with a brief discussion of limitations and possible extensions.

\section{Context}
\subsection{Notations}
Given two measurable spaces $(\spaceX_1, \mathcal{A}_1)$ and $(\spaceX_2, \mathcal{A}_2)$,
we define an unnormalized kernel (resp. a Markov kernel) $K$ from $\spaceX_1$ to $\spaceX_2$
as a function $\spaceX_1 \times \mathcal{A}_2 \to [0, \infty]$ such that
for any set $A \in \mathcal{A}_2$ the function $x \in \spaceX_1 \mapsto K(x, A)$ is measurable
and for any $x \in \spaceX_1$ the function $A \in \mathcal{A}_2 \mapsto K(x, A)$
is a finite measure (resp. a probability measure).
Given some positive measurable function $f\colon \spaceX_1 \times \spaceX_2 \to [0, \infty]$,
we define the function $K(f)\colon \spaceX_1 \to [0, \infty]$ as follows:
\begin{equation*}
  K(f)(x_0) = \int f(x_0, x) K(x_0, dx)
  .
\end{equation*}
Similarly, given some measure $\K(dx)$ and some positive measurable function $f\colon \spaceX \to [0, \infty]$
we define the quantity $\K(f) \in [0, \infty]$ as follows:
\begin{equation*}
  \K(f) = \int f(x) \K(dx)
  .
\end{equation*}

The notation $t:s$ denotes the sequence $(t, t+1, \dots, s)$,
$x_{t:s}$ denotes a sequence $(x_t, \dots, x_s)$, $x^{n:m}$ denotes a sequence $(x^n, \dots, x^m)$
and $x_{t:s}^{n:m}$ denotes the collection of all $x_i^j$ with $t \leq i \leq s$ and $n \leq j \leq m$,
For spaces, $\spaceX_{t:s}$ denotes the product space $\spaceX_t \times \dots \spaceX_s$.
Given a sequence of kernels $(K_t)_{t\geq 2}$ from $\spaceX_{t-1}$ to $\spaceX_t$
and a measure $\K_1$ on $\spaceX_1$, $K_{t:s}$ denotes the joint kernel
from $\spaceX_{t-1}$ to $\spaceX_{t:s}$ as follows:
\begin{equation*}
  K_{t:s}(x_{t-1}, dx_{t:s}) = K_t(x_{t-1}, dx_t) \dots K_s(x_{s-1}, dx_s)
\end{equation*}
and the notation $\K_{1:t}$ denote the joint measure on $\spaceX_{1:t}$ as follows:
\begin{equation*}
  \K_{1:t}(dx_{1:t}) = \K_1(dx_1) \dots K_t(x_{t-1}, dx_t)
  .
\end{equation*}
Given a sequence of positive measurable functions $(G_t)_{t\geq 2}$ on $\spaceX_{t-1} \times \spaceX_t$
and $G_1$ on $\spaceX_1$,
$G_{t:s}$ (resp. $G_{1:t}$) denotes the product function on $\spaceX_{t-1:s}$ (resp $\spaceX_{1:t}$):
\begin{align*}
  G_{t:s}(x_{t-1:s}) & = G_t(x_{t-1}, x_t) \dots G_s(x_{s-1}, x_s)
  ,
  \\
  G_{1:s}(x_{1:t}) & = G_1(x_1) \dots G_t(x_{t-1}, x_t)
  .
\end{align*}

For two kernels $K$ and $U$ from $\spaceX_1$ to $\spaceX_2$ such that, for each $x_1\in \spaceX_1$,
the measure $K(x_1, dx_2)$ is absolutely continuous with respect to $U(x_1, dx_2)$,
we denote by $\frac{dK}{dU}$ the Radon-Nikodym derivative between these two measures:
\begin{equation*}
  \frac{dK}{dU}(x_1, x_2) = \frac{K(x_1, dx_2)}{U(x_1, dx_2)}(x_2)
\end{equation*}

We denote by $\one$ any constant unit function.
We denote by $\mathcal N(m,\Sigma)$ or $\mathcal N(dx, m,\Sigma)$
the multivariate Gaussian distribution with mean $m$ and covariance matrix $\Sigma$.
We denote $\mathrm{Categorical}(w_1, \dots, w_n)$
the categorical distribution that gives probability $w_i/\sum_{j=1}^n w_j$ to $i \in 1:n$.

\subsection{State-space models}

Our main motivation comes from the problem of performing inference for state-space models (SSMs),
which are a class of models that link a sequence of observations $y_{1:T}$
to a latent variable $x_{1:T}$ that takes values in $\spaceX$
and typically follows a Markovian dependency structure.

The first component of an SSM is the emission density $\pi(y_t \mid x_t)$
(with respect to, typically, the Lebesgue measure) of an individual observation given the latent
value at time $t$. The second component is a latent Markov chain, with an
initial distribution denoted by $\pi(dx_1)$, and Markov transition kernels
denoted by $\pi(dx_t \mid x_{t-1})$. In an SSM, the standard tasks are to infer
the filtering distributions:
\begin{equation*}
  \pi(dx_t \mid y_{1:t}), \quad t \in 1:T
\end{equation*}
the smoothing distribution:
\begin{equation*}
  \pi(dx_{1:T} \mid y_{1:T})
\end{equation*}
and the marginal likelihood:
\begin{equation*}
  \pi(y_{1:T}) .
\end{equation*}
In general, closed-form expressions for these tasks are not available, and the
typical solution is to construct Monte Carlo estimates of these targets.

Obtaining the smoothing distribution is an inference problem on a latent space $\spaceX^T$,
which can be very high-dimensional,
but there is a lower intrinsic dimension,
thanks to the sequential dependency structure of the latent space.
To exploit this sequential dependency,
we define a sequence of growing target probability measures $\nu_t$
on the space $\spaceX^t$ as:
\begin{equation*}
  \nu_t(dx_{1:t}) = \pi(dx_{1:t} \mid y_{1:t})
  .
\end{equation*}
The smoothing distribution corresponds to the terminal target $\nu_T$
and the filtering distributions correspond to marginals of each $\nu_t$.

To infer our targets $\nu_t$ sequentially (from $\nu_1$ to $\nu_T$), Bayes' rule gives the following relations:
\begin{gather}
  \pi(dx_1 \mid y_1)
  = \frac{\pi(y_1\mid x_1) \pi(dx_1)}{\pi(y_1)}
  \label{eq:ssm_fkm_init}
  \\
  \begin{aligned}
    & \pi(dx_{1:t+1} \mid y_{1:t+1})
    \\
    & = \frac{\pi(y_{t+1}\mid x_{t+1})\pi(dx_{1:t+1} \mid y_{1:t})}{\pi(y_{t+1}\mid y_{1:t})}
    \\
    & = \frac{\pi(y_{t+1}\mid x_{t+1})\pi(dx_{t+1}\mid x_t)\pi(dx_{1:t} \mid y_{1:t})}
    {\pi(y_{t+1}\mid y_{1:t})}
  \end{aligned}
  \label{eq:ssm_fkm}
\end{gather}
which can be seen as a recursive relation between $\nu_{t+1}$ and $\nu_t$.

Let us assume that we know how to sample from the transition kernels $\pi(dx_{t+1} \mid x_t)$
(as well as form the initial distribution $\pi(dx_1)$)
and that we can compute the emission densities $\pi(y_t\mid x_t)$.
In this setting, equations \eqref{eq:ssm_fkm_init} and \eqref{eq:ssm_fkm} describe the sequential construction of $\nu_{1:T}$ and define a Feynman--Kac model.
We will detail this notion in the following sections and explain
how it can be used to construct importance sampling approximations of the targets $\nu_{1:T}$
within a sequential Monte Carlo algorithm.

\subsection{The Feynman-Kac formalism}

We are interested in approximating the targets $\nu_{1:T}$ sequentially.
To this end, we begin by introducing Feynman-Kac models \citep{del_moral_feynman-kac_2004},
which provide a useful framework for describing sequentially constructed measures and sequential Monte Carlo algorithms.

A Feynman-Kac model $(\M_{1:T}, G_{1:T})$ up to time $T$ is composed of
a Markov chain $\M_{1:T}$ from which we can sample, and
non-negative weight functions $G_t(x_{t-1}, x_t)$ for $t \in 1:T$.

We define the target path measures $\nu_{1:T}$ and the normalizing constants $Z_t$ associated with a Feynman-Kac model as:
\begin{align*}
  \nu_t(dx_{1:t}) & = \frac{1}{Z_t} G_{1:t}(x_{1:t}) \M_{1:t}(dx_{1:t})
  ,                                                                     \\
  Z_t & = \int G_{1:t}(x_{1:t}) \M_{1:t}(dx_{1:t})
  .
\end{align*}
These measures satisfy a relation similar to \eqref{eq:ssm_fkm}:
\begin{gather}
  \nu_{t+1} = \frac{Z_t}{Z_{t+1}} G_{t+1} \nu_{t} M_{t+1}
  .\label{eq:rec_fkm}
\end{gather}
For an SSM, this leads to a canonical Feynman-Kac model $(\M_{1:t}, G_{1:t})$ defined as:
\begin{align}
  \begin{split}
    \M_1(dx_1) & = \pi(dx_1)
    \\
    M_t(x_{t-1}, dx_t) & = \pi(dx_t \mid x_{t-1})
    \\
    G_t(x_{t-1}, x_t) & = \pi(y_t \mid x_t)
    .
  \end{split}
  \label{eq:fkm_canon}
\end{align}
With this definition, the target path measures coincide with the targets defined in the previous section,
that is, $\nu_t(dx_{1:t}) = \pi(dx_{1:t} \mid y_{1:t})$
and the normalizing constants correspond to the marginal likelihoods $Z_t = \pi(y_{1:t})$.

For a given Feynman-Kac model, we define the cost-to-go function $H_{t\to s}$ (from $t$ to $s$) as:
\begin{equation}
  H_{t\to s}(x_t) = M_{t+1:s}(G_{t+1:s})(x_t)
  \label{eq:cost-to-go}
\end{equation}
which is, up to a multiplicative constant, the Radon-Nikodym derivative
between the target path measure $\nu_s$ and the marginal of $\nu_t$ over $x_{1:t}$.
By convention, we set $H_{s \to s} \equiv 1$ and $H_{0\to t} \equiv Z_t$.

For the canonical Feynman-Kac model associated with an SSM,
a cost-to-go function is the likelihood of future observations given the current latent state:
$H_{t\to s}(x_t) = \pi(y_{t+1:s}\mid~x_t)$.

\subsection{Sequential Monte Carlo algorithms}

A sequential Monte Carlo (SMC) algorithm is a method for sampling from
a sequence of target measures $\nu_1, \dots, \nu_T$, known up to normalizing constants $Z_1, \dots, Z_T$.

Given a Feynman-Kac model $(\M_{1:T}, G_{1:T})$,
since each target $\nu_t$ is available in closed form from the previous target $\nu_{t-1}$ (see \eqref{eq:rec_fkm}),
we can construct Monte Carlo approximations sequentially, from $\nu_1$ to $\nu_T$.

A standard sequential Monte Carlo algorithm (see \autoref{algo:SMC}) follows the evolution of $N$ particles $x_t^{1:N}$ with associated weights $w_t^{1:N}$
over discrete time $t$, starting at $1$ and ending at $T$.
These particles are sampled and weighted according to the Feynman-Kac model
to construct an importance sampling approximation
of the targets $\nu_t(dx_{1:t})$.
We refer to $\M_{1:T}$ as the proposal Markov chain and $G_{1:T}$ the weight functions.

If we were to reduce the algorithm to a sequence of importance sampling steps (i.e.,\ without resampling),
we would observe particle degeneracy, that is, a few particles would receive most of the weight,
resulting in a poor approximation.
One of the key features of an SMC algorithm that mitigates particle degeneracy
is the resampling step: particles with low weights are discarded, while particles with large weights are selected to produce offspring.
Formally, this can be described as selecting ancestors indices $a_t^{1:N} \in \N$
for the next generation of $N$ particles at $t+1$, according to the weights $w_t^{1:N}$.

To simplify notation, when the context creates no ambiguity,
we denote by $\xi_{1:t}^n$ the trajectory of a particle, taking into account its ancestors:
\begin{equation*}
  \xi_{1:t}^n = (\xi_{1:t-1}^{a_{t-1}^n}, x_t^n)
  .
\end{equation*}

From the output ($x_{1:T}^{1:N}$, $a_{1:T}^{1:N}$, $w_{1:T}^{1:N}$) of an SMC algorithm, we construct
the following estimators of $\nu_t(\varphi)$, for a $\R$-valued function $\varphi$, and of $Z_t$:
\begin{align}
  \hat \nu_t^N(\varphi) & = \frac 1 {\sum_{n=1}^N w_t^n} \sum_{n=1}^N w_t^n \varphi(\xi_{1:t}^n)
  \\
  \hat Z_t^N & = \prod_{s=1}^t \frac 1 N \sum_{n=1}^N w_s^n = \hat Z_{t-1}^N \frac 1 N \sum_{n=1}^N w_t^n
  \label{eq:defZhat}
  .
\end{align}

When the targets $\nu_t$ correspond to the previously introduced
inference targets $\pi(dx_{1:t}\mid y_{1:t})$ of an SSM,
the resulting SMC algorithm is called a particle filter.
When the Feynman-Kac model is the canonical Feynman-Kac model of an SSM defined by \eqref{eq:fkm_canon},
an SMC algorithm is called a bootstrap particle filter.
The bootstrap particle filter is the simplest particle filter to implement,
and it will serve as a baseline reference algorithm throughout this article.

\begin{algorithm}
  \caption{Standard Sequential Monte Carlo}
  \label{algo:SMC}
  \KwIn{A Feynman-Kac model $(\M_{1:T}, G_{1:T})$, a number of particles $N$.}
  \KwOut{Particles $x_{1:T}^{1:N}$, weights $w_{1:T}^{1:N}$, ancestors index $a_{1:T}^{1:N}$.}
  \BlankLine
  \AtPos{$t=1$}{
    \Sample{} $x_1^n \sim \M_1(d x_1)$ for $n\in 1:N$. \;
    \Compute{} weights $w_1^n = G_1(x_1^n)$ for $n\in 1:N$. \;
    \Resample{} to get ancestors $a_1^n \sim \mathrm{Categorical}(w_1^{1:N})$ for $n\in 1:N$. \;
  }
  \AtPos{$t\in 2:T$}{
    \Sample{$x_t^n \sim M_t(x_{t-1}^{a_{t-1}^n}, d x_t)$} for $n \in 1:N$. \;
    \Compute{} weights $w_t^n = G_t(x_{t-1}^{a_{t-1}^n}, x_t^n)$ for $n \in 1:N$. \;
    \Resample{} to get ancestors $a_t^n \sim \mathrm{Categorical}(w_t^{1:N})$ for $n\in 1:N$. \;
  }
\end{algorithm}

\subsection{Objectives}

The bootstrap particle filter is simple but often leads to inefficient algorithms due to particle degeneracy,
arising from the use of $\pi(dx_t \mid x_{t-1})$ as the proposal kernels $M_t(x_{t-1}, dx_t)$.
The goal of our work is to automatically refine the bootstrap particle filter by proposing a method to iteratively construct
better proposals, that is, better Feynman-Kac models.

More precisely, given a reference Feynman-Kac model $(\refer \M_{1:T}, \refer G_{1:T})$,
the bootstrap filter in our case, we are interested
in accurately estimating the full target $\refer \nu_T$ and its normalizing constant $\refer Z_T$.
Within the constraint of preserving these two, we will try to find the best Feynman-Kac model.

\section{Method}
\subsection{Auxiliary Feynman-Kac models with preserved final target}

Our goal is to develop a method for constructing optimal Feynman-Kac models that sample from the full target distribution.
We introduce auxiliary Feynman-Kac models, inspired by the auxiliary particle filter
(\citealt{pitt_filtering_1999}, see also Chap. 10 of
\citealt{chopin_introduction_2020}).
They are defined by a proposal chain $\M_{1:T}$
and auxiliary weight functions $\eta_{1:T}$, with $\eta_0 \equiv \eta_T \equiv 1$,
such that the auxiliary Feynman-Kac model $(\M_{1:T}, \eta_{1:T})$
given a reference model $(\refer \M_{1:T}, \refer G_{1:T})$,
defines a Feynman-Kac model $(\M_{1:T}, G_{1:T})$
that satisfies:
\begin{equation}
  \begin{split}
    G_1(x_1) & = \refer G_1(x_1) \frac{\refer \M_1(dx_1)}{\M_1(dx_1)} \eta_1(x_1)
    \\
    G_t(x_{t-1}, x_t) & = \refer G_t(x_{t-1}, x_t)
    \\
    & \times \frac{\refer M_t(x_{t-1}, dx_t)}{ M_t(x_{t-1}, dx_t)} \frac{\eta_t(x_t)}{\eta_{t-1}(x_{t-1})}
    .
  \end{split}
  \label{eq:apffkm}
\end{equation}

In this context, at time $t$, the target path measure $\nu_t$ is given by:
\begin{equation}
  \nu_t = \frac{G_{1:t}}{\int G_{1:t} d \M_{1:t}} \M_{1:t} = \frac{\eta_t}{\int \eta_t d\refer \nu_t} \refer \nu_t
  \label{eq:afkm_path}
\end{equation}
the normalizing constant $Z_t$ is:
\begin{equation*}
  Z_t = \int G_{1:t}d\M_{1:t} = \refer Z_t\int \eta_t d\refer \nu_t
\end{equation*}
and the cost-to-go function $H_{t\to T}$ is:
\begin{equation*}
  H_{t\to T} = M_{t+1:T}(G_{t+1:T}) = \frac{\refer H_{t\to T}}{\eta_t}
  .
\end{equation*}

Since we require that $\eta_T \equiv 1$ we have $\nu_T = \refer \nu_T$
and $Z_T = \refer Z_T$.
The above relations describe a family of  Feynman-Kac models with the same full
target and final normalizing constant.
In fact, we describe all the Feynman-Kac models satisfying these two properties.
If we set $\M_{1:T} \equiv \refer M_{1:T}$ and $\eta_{1:T-1} \equiv 1$, we recover exactly the reference.
In practice, as shown by \eqref{eq:afkm_path}, the auxiliary weights allow us to use a different sequence of measures
that still lead to $\nu_T$.

In the context of an SSM, it is crucial to emphasize
that an SMC algorithm derived from an auxiliary Feynman-Kac model will
not target the usual particle filtering distributions,
since the intermediate targets are modified,
although it remains possible to produce filtering estimates.
Therefore, we do not expect to obtain optimal algorithms for the filtering task.
Instead, our objective is to achieve optimality for the marginal likelihood,
that is, for the estimator of $Z_T$,
and, since we preserve the final target, this approach
should also be efficient for the smoothing task.

\subsection{Variance of the marginal likelihood estimator}

We denote by $\hat Z_T^N$ the Monte Carlo estimator of the normalizing constant $Z_T$
obtained from the output of an auxiliary Feynman-Kac model and defined in \eqref{eq:defZhat}.

This estimator is unbiased, $\expect{}[\hat Z_T^N] = Z_T$,
and its variance satisfies \citep{doucet_tutorial_2011}:
\begin{equation}
  \frac{\var{}[\hat Z_T^N]}{Z_T^2}
  = \frac 1 N \sum_{t=1}^T \bigg(
    \\
    \int G_t H_{t\to T} d\nu_T\int \frac{1}{G_t H_{t\to T}}d\nu_T - 1
  \bigg)
  \\
  + o\left(\frac 1 N\right)
  \label{eq:asvar}
  .
\end{equation}

The first integral term in~\eqref{eq:asvar} can be expressed, for an auxiliary Feynman-Kac model, as:
\begin{equation}
  G_t(x_{t-1}, x_t) H_{t\to T}(x_t)
  \\
  = \frac{\refer G_t(x_{t-1}, x_t) \refer H_{t\to T}(x_t)}{\eta_{t-1}(x_{t-1})}
  \frac{\refer M_t(x_{t-1}, dx_t)}{M_t(x_{t-1}, dx_t)}
  \label{eq:varterm}
  .
\end{equation}

The variance of this estimator is of particular interest for various reasons.
First, it determines the performance of the popular PMMH algorithm
\citep{Andrieu2010},
which may be used to sample from the posterior distribution of
the parameters of the considered state-space model.
Second, a low variance also implies reduced weight degeneracy and
suggests that the full proposal distribution is close to the full target distribution.
For this reason, we use this variance to define optimality in the next section.

The summand in~\eqref{eq:asvar} is a (reverse) $\chi^2$ divergence between the target $\nu_T$ and
the proposal ($\frac 1 {G_t H_{t\to T}} \nu_T$ up to a constant).
More precisely, given two probability measures $\nu$ and $\mu$ on $\spaceX$,
and a non-negative function $f$ on $\spaceX$,
we define a chi-squared divergence between $\nu$ and $\mu$, as follows:
\begin{equation*}
  \chitwo(\nu\mid\mu) = \int \frac {d\nu}{d\mu}d\mu \int \frac{d\mu}{d\nu} d\mu - 1
\end{equation*}
and to simplify notation,
we define a chi-squared divergence of $f$ relative to $\mu$
as the chi-squared divergence between $\mu$ and $\nu \propto f\mu$,
as follows:
\begin{equation*}
  \chitwo[\mu](f) = \int \frac 1 f d\mu \int f d\mu - 1
  .
\end{equation*}
Then, using~\eqref{eq:varterm}, \eqref{eq:asvar} may be rewritten as:
\begin{equation}
  \frac{\var{}[\hat Z_T^N]}{Z_T^2}
  = \frac 1 N \sum_{t=1}^T
  \chitwo[\nu_T]\left(
    \frac{\refer G_t \refer H_{t\to T}}{\eta_{t-1}}
    \frac{d\refer M_t}{dM_t}
  \right)
  \label{eq:varchitwo}
  \\
  + O\left(\frac 1 {N^2}\right)
  .
\end{equation}
See \autoref{ann:variance} for further details on how to derive \eqref{eq:varchitwo}
from \citet{doucet_tutorial_2011}.

In our context, we aim to minimize the leading term in~\eqref{eq:varchitwo}
with respect to $\eta_{t-1} M_t$, while keeping $\refer G_{1:T}$ and $\refer M_{1:T}$ fixed.

Therefore, to simplify the expressions, we introduce the following quantities:
\begin{equation*}
  \U_1 = \eta_0 \M_1 \quad U_t = \eta_{t-1} M_t
\end{equation*}
that is,\ $\U_1$ is a finite measure and the $U_t$'s are unnormalized kernels.
We can recover both $\eta_{t-1}$ and $M_t$ from $U_t$ via:
\begin{align*}
  \eta_{t-1}(x_{t-1}) & = \int U_t(x_{t-1}, dx_t)
  ,                                                                        \\
  M_t(x_{t-1}, dx_t) & = \frac{1}{\eta_{t-1}(x_{t-1})} U_t(x_{t-1}, dx_t)
  .
\end{align*}

The auxiliary Feynman-Kac model notations allow, via \eqref{eq:varchitwo},
to precisely characterize which quantity, which discrepancy, and which measure
should drive our learning algorithms, given that we want to minimize this variance.
The main consequence of this formulation is that optimality can be understood from \eqref{eq:varchitwo}
on the unconstrained set of unnormalized measures $U_{1:T}$
more directly than from \eqref{eq:asvar}, which is defined on the set of Feynman-Kac models
constrained by their full target distribution.

\subsection{Optimality}

\subsubsection{Global optimality}
We define optimality in our context as finding $U_t$ (or equivalently $M_t$ and $\eta_t$)
that minimizes the leading term of~\eqref{eq:asvar}.
The minimizer $U_t^\star$ follows directly from~\eqref{eq:varchitwo}
and can be written as:
\begin{equation}
  \begin{split}
    \U_1^\star & = \refer G_1 \refer H_{1\to T} \refer M_1
    ,                                                      \\
    U_t^\star & = \refer G_t \refer H_{t\to T} \refer M_t
    ,\quad t \geq 2
    .
  \end{split}
  \label{eq:globopt}
\end{equation}
By integration, the normalizing constant is,
by definition of the cost-to-go function \eqref{eq:cost-to-go}:
\begin{equation}
  \eta_{t-1}^\star = \refer M_t(\refer G_t \refer H_{t\to T}) = \refer H_{t-1\to T}
  \label{eq:globopt_aux}
\end{equation}
This leads to the following optimal proposal kernels:
\begin{align*}
  \M_1^\star & = \frac{\refer G_1 \refer H_{1\to T}}{Z_T}\refer \M_1
  ,                                                                                 \\
  M_t^\star & = \frac{\refer G_t \refer H_{t\to T}}{\refer H_{t-1\to T}}\refer M_t
  ,\quad t\geq 2
  .
\end{align*}

If the reference is the bootstrap particle filter, this yields:
\begin{align*}
  M_1^\star(dx_1) & = \pi(dx_1 \mid y_{1:T})
  \\
  M_t^\star(x_{t-1}, dx_t) & = \pi(dx_t \mid x_{t-1}, y_{t:T})
  \\
  \eta_{t-1}^\star(x_{t-1}) & = \pi(y_{t:T} \mid x_{t-1})
  .
\end{align*}
This corresponds to sampling directly from the full target distribution,
i.e.,\ the smoother in this context \citep{guarniero_iterated_2017, heng_controlled_2020}:
\begin{equation*}
  \nu_t^\star(dx_{1:t}) = \pi(dx_{1:t} \mid y_{1:T}).
\end{equation*}

Recast in our formalism, the approaches of \citet{guarniero_iterated_2017} and \citet{heng_controlled_2020}
approximate the optimal sequence $\U_{1:T}$
by means of the following explicit backward recursion, for $t=T-1, T-2, \dots, 1$:
\begin{equation}
  U_{t}^\star = \refer G_t U_{t+1}^\star(\one) \refer M_t
  \label{eq:backschm}
\end{equation}
which follows from \eqref{eq:globopt} and \eqref{eq:globopt_aux}.
This implies the following relations for $\eta_{t-1}^\star$ and $M_t^\star$:
\begin{align*}
  \M_1^\star & = \frac{\refer G_1 \eta_1^\star}{\int \refer G_1\eta_1^\star d\refer \M_1} \refer \M_1
  \\
  M_t^\star & = \frac{\refer G_t \eta_t^\star}{\refer M_t(\refer G_t \eta_t^\star)}\refer M_t
  \\
  \eta_{t-1}^\star & = \refer M_t(\refer G_t \eta_t^\star)
  .
\end{align*}
If the reference is the bootstrap particle filter,
we recover the backward smoothing recursion \citep{doucet_tutorial_2011}:
\begin{equation*}
  \pi(x_t \mid x_{t-1}, y_{t:T})
  \\
  = \frac{
    \pi(y_t \mid x_t)\pi(y_{t+1:T}\mid x_t)\pi(x_t \mid x_{t-1})
  }{\pi(y_{t:T}\mid x_{t-1})}
  .
\end{equation*}

\subsubsection{Tractability issues and local optimality}

Global optimality introduces some computational difficulties.
The main difficulty is that this optimum is typically intractable,
since the optimal target at time $t$, $\nu_t^\star$,
is precisely the marginal of the full target $\nu_T^\star = \nu_T$.
This can be addressed by building approximations iteratively, alternating
forward and backward passes over the data, as in
\citet{guarniero_iterated_2017} and \citet{heng_controlled_2020}.
However, as we shall observe in our experiments,
this approach requires a sufficiently good initialization (i.e.,\ the first
forward pass), which is non-trivial in cases where the reference Feynman-Kac model
leads to a very inefficient particle filter (e.g., a bootstrap filter for a
state-space model with highly informative data).

Instead, our method relies on the concept of local optimality, which can be intuitively understood
as optimality with respect to a finite horizon of $L$ steps ahead
($\refer H_{t\to t+L}$ in \eqref{eq:globopt}),
rather than with respect to the full remaining sequence $\refer H_{t\to T}$.

This allows us to follow a forward scheme,
which can be used either as an initial proposal for backward-scheme algorithms
or to construct an online algorithm — something not achievable with a purely backward scheme.

More precisely, local optimality is defined in the same sense as
the local optimality for a guided particle filter, described in \citet{doucet_sequential_2000}.
The main idea underlying this notion of optimality is to ignore knowledge of the full target and only
include weight functions up to time $t$.
Namely, we aim to optimize the variance of the estimator $\hat Z_t^N$ of $Z_t$ instead of $\hat Z_T^N$ of $Z_T$.

We look for an optimum with the target path $\nu_{1:T}$ fixed (i.e.,\ at $\eta_{1:T}$ fixed).
By reducing this problem to minimizing the variance of the importance weights $w_t^n$,
it is well known \citep{doucet_sequential_2000} that this leads to the following optimum:
\begin{equation}
  \begin{split}
    \M_1^{\mathrm{loc}} & = \frac{\refer G_1 \eta_1}{\int \refer G_1\eta_1 d\refer\M_1} \refer\M_1
    \\
    M_t^{\mathrm{loc}} & = \frac{\refer G_t \eta_t}{\refer M_t(\refer G_t \eta_t)} \refer M_t
    .
  \end{split}
  \label{eq:locopt}
\end{equation}

Local optimality provides insight into how to construct the best proposals for a fixed target path $\nu_{1:T}$.
In the next section, we develop a scheme that connects this notion of local optimality to
a global optimal path.

\subsection{Iterated forward scheme}

The optimal auxiliary weight functions $\eta_t^\star = \refer H_{t\to T}$
correspond to the cost-to-go from time $t$ to final time $T$.
Therefore, if we fix a number $L$ of future time points to take into account
when constructing our proposals,
we can define auxiliary functions $\eta_{0:T}^{(L)}$ as
the cost-to-go from time $t$ to time $t+L$:
\begin{equation*}
  \eta_t^{(L)} =
  \begin{cases}
    \refer H_{t \to t+L} & \text{if $t+L \leq T$}
    \\
    \refer H_{t\to T} = \eta_t^\star & \text{else.}
  \end{cases}
\end{equation*}
These auxiliary functions converge to $\eta_t^\star$ as $L$ increases.

When the bootstrap particle filter is used as a reference, $\eta_t^{(L)}$ satisfies
\begin{align*}
  & \eta_t^{(L)}(x_t)
  \\&= \int \refer G_{t+1:t+L}(x_{t:t+L}) \refer M_{t+1:t+L}(x_t, dx_{t+1:t+L})
  \\&= \int \prod_{s=t+1}^{t+L} \pi(y_{t+s} \mid x_{t+s}) \pi(dx_{t+s} \mid x_{t+s-1})
  \\&= \pi(y_{t+1:t+L}\mid x_t)
\end{align*}
for $t+L \leq T$, so that $\eta_t^{(L)}$ weights $x_t$ according to the information contained in $L$ future observations.

This establishes a connection with the fixed-lag SMC literature \citep{doucet_fixed-lag_2004},
which focuses on sampling from $\pi(dx_{1:t}\mid y_{1:t+L})$,
which is precisely the target $\nu_t^{(L)}$ associated with the auxiliary weight $\eta_t^{(L)}$.
Rapid convergence of this target toward the smoothing distribution was shown by \citet{olsson_sequential_2008},
and by extension, the target $\nu_t^{(L)}$ should rapidly approximate
the target induced by the backward scheme.

There are two main differences between our approach and fixed-lag SMC.
First, instead of sampling blocks $x_{t:t+L}$ to estimate $\nu_t^{(L)}$,
we sample from an approximation of its $x_t$-marginal.
Second, we build our approximation iteratively over $L$.
As a result, our algorithm uses an increasing value of $L$,
compared to the standard block sampling method, which uses a fixed $L$.

These auxiliary weights may be computed recursively, using a backward scheme:
\begin{equation*}
  \eta_t^{(L+1)} = \refer M_{t+1}(\refer G_{t+1} \eta_{t+1}^{(L)})
\end{equation*}
starting from $\eta_t^{(0)} \equiv 1$, corresponding to the reference.
We recall that $\eta_T \equiv 1$ is required to preserve the last target.
One crucial difference compared to the backward scheme is that we only need to integrate up to $L$ times
to compute the auxiliary weight as opposed to $T$ times under global optimality.

By applying local optimality \eqref{eq:locopt} to fixed auxiliary weights $\eta_t^{(L)}$,
we obtain an auxiliary model $(\M_{1:T}^{(L+1)}, \eta_{0:T}^{(L)})$ defined by:
\begin{equation*}
  \begin{split}
    \M_1^{(L+1)} & = \frac{\refer G_1 \eta_1^{(L)}}{\int \refer G_1 \eta_1^{(L)} d\refer \M_1} \refer\M_1
    \\
    M_t^{(L+1)} & = \frac{\refer G_t \eta_t^{(L)}}{\refer M_t(\refer G_t\eta_t^{(L)})} \refer M_t
    .
  \end{split}
\end{equation*}
This leads to the following relation, similar to~\eqref{eq:backschm}:
\begin{equation}
  U_t^{(L+1)} = \refer G_t U_{t+1}^{(L)}(\one) \refer M_t
  \label{eq:fwdschm}
  .
\end{equation}

Although this equation remains a backward recursion with respect to time $t$, it can
also be interpreted as a forward recursion with respect to $L$.
We use this latter interpretation to derive our practical, forward-only algorithm, as
explained in the next section.

\section{Implementation}

\subsection{Parametrization using twisted Feynman-Kac models}

We use the same family of approximating proposal kernels  as in
\citet{guarniero_iterated_2017} and
\citet{heng_controlled_2020}, that is, we
parametrize the unnormalized kernel in terms of a twisting function
$\varphi_t$:
\begin{equation*}
  U_t(x_{t-1}, dx_t) = \varphi_t(x_{t-1}, x_t) \refer M_t(x_{t-1}, dx_t)
  .
\end{equation*}
This choice is motivated by the fact that the optimal kernel satisfies
\begin{equation*}
  U_t^\star(x_{t-1}, dx_t)
  \\
  = \refer G_t(x_{t-1}, x_t) \eta_t^\star(x_t) \refer M_t(x_{t-1}, dx_t)
\end{equation*}
according to~\eqref{eq:globopt}.
Hence, the problem reduces to learning a strictly positive function $\varphi_t$ of the form:
\begin{align*}
  \varphi_t(x_{t-1}, x_t)
  & = \frac{U_t(x_{t-1}, dx_t)}{\refer M_t(x_{t-1}, dx_t)}
  \\
  & \approx \refer G_t(x_{t-1}, x_t) \eta_t^\star(x_t)
  .
\end{align*}
When the reference is derived from the bootstrap particle filter,
$\refer G_t$ depends only on $x_t$, and is constant with respect to $x_{t-1}$.
In that case, we restrict $\varphi_t$ to be constant in $x_{t-1}$.

This parametrization is especially useful when the transition kernels admit
conjugacy properties, e.g., when $\refer M_t(x_{t-1}, dx_t)$ is
Gaussian (conditional on $x_{t-1}$).  In that
case, if $\varphi_t$ is chosen in the space of log-quadratic functions, that is,
provided
\begin{align*}
  \M_1(dx_1) & = \mathcal{N}(m_1, \Sigma_1)
  \\
  M_t(x_{t-1}, dx_t) & = \mathcal{N}\left(m_t(x_{t-1}), \Sigma_t(x_{t-1})\right)
\end{align*}
and
\begin{equation}
  \varphi_t(x_t) = \exp\left(- \frac 1 2  x_t'A_t x_t - x_t' b_t - \frac 1 2 c_t\right)
  \label{eq:logquad}
\end{equation}
under the condition $\Sigma_t(x_{t-1}) + A_t\gg  0$ for all $t$, all quantities of interest are available in closed form:
\begin{gather*}
  M_t = \frac{1}{U_t(\one)} U_t = \frac{\varphi_t}{\refer M_t(\varphi_t)} \refer M_t
  \\
  \eta_t = U_t(\one) = \refer M_t(\varphi_t)
  .
\end{gather*}

Function $\varphi_t$ with nice conjugacy properties seems to be typical of Gaussian applications:
we are not aware of non-Gaussian examples that admit such functions.
The methods presented below use this parametrization, but could be adapted
to optimize over a class of unnormalized kernels $U_t$ instead of $\varphi_t$ when necessary.

\subsection{Recursively computing the twisting functions}

\begin{algorithm}
  \caption{Forward SMC Training}
  \label{algo:fwd_th}
  \KwIn{
    A reference Feynman-Kac model $(\refer \M_{1:T}, \refer G_{1:T})$,
    a number of particles $N$, a number of iterations $L^{\max}$
    and sets of strictly positive functions $\Psi_{1:T}$.
  }
  \KwOut{Approximations $\hat \varphi_{1:t}^{(L)}$ for $1\leq L \leq L^{\max}$ of the forward scheme.}
  \BlankLine
  \Set $\hat\varphi_{1:T}^{(0)} \equiv 1$,
  $\hat \M_1^{(0)}\sim\refer \M_1$, $\hat M_{2:T}^{(0)}\equiv \refer M_{2:T}$ and $\hat \eta_{1:T}^{(0)}\equiv 1$.\;
  \For{$L \in 0:L^{\max}-1$}{
    \For{$t \in 1:T$}{
      \Approximate $\refer G_t \hat\eta_t^{(L)}$ by $\hat \varphi_t^{(L+1)} \in \Psi_t$.\;
      \Set $\hat\eta_{t-1}^{(L+1)} = \refer M_t(\hat\varphi_t^{(L+1)})$.\;
      \Set $\hat M_t^{(L+1)} = \frac{\hat\varphi_t^{(L+1)}}{\hat\eta_{t-1}^{(L+1)}}\refer M_t$.\;
      \tcp{Sample particles according to trained proposal}
      \Sample and \Compute $x_{t}^{(L+1),n}$, $w_{t}^{(L+1),n}$ and $a_t^{(L+1), n}$
      according to the $t$-th SMC step (\autoref{algo:SMC})
      with $M_t = \hat M_{t}^{(L+1)}$, $\eta_{t-1:t} = \eta_{t-1:t}^{(L)}$
      and ancestor $\ancestor x_{t-1}^{(L+1),n}$.\;
      \Denote $\ancestor x_{t}^{(L+1), n} = x_{t}^{(L+1),a_{t}^{(L+1), n}}$. \;
    }
  }
\end{algorithm}

The main idea in approximating the schemes
is to substitute approximations $\hat\varphi_t$ (resp. $\hat \varphi_t^{(L)}$)
of the optimal twisting functions
$\varphi^\star_t=dU_t^\star/d\refer M_t$ (resp. $\varphi^{(L)}_t=dU_t^{(L)}/d\refer M_t$)
into the recursions \eqref{eq:backschm} (resp. \eqref{eq:fwdschm}).
Several methods exist to construct these approximations. In this section, we detail an adaptation to the forward scheme \eqref{eq:fwdschm}
of the approach proposed by \citet{heng_controlled_2020}
for the backward scheme \eqref{eq:backschm}.
See \autoref{ann:csmc} for more details on the backward implementation of \citet{heng_controlled_2020}.

Expressing the forward scheme \eqref{eq:fwdschm} in terms of the twist functions $\varphi^{(L)}_t=dU_t^{(L)}/d\refer M_t$,
we obtain:
\begin{equation*}
  \varphi_t^{(L+1)} \refer M_t = \refer G_t \refer M_{t+1}(\varphi_{t+1}^{(L)}) \refer M_t
  .
\end{equation*}
This gives the following recursion:
\begin{equation}
  \varphi_t^{(L+1)} = \refer G_t \refer M_{t+1}(\varphi_{t+1}^{(L)})
  \label{eq:fwdschm_twist}
  .
\end{equation}
This equation expresses the scheme entirely in terms of the twist functions $\varphi_t$.
Following \eqref{eq:fwdschm_twist}, we iteratively construct $\hat \varphi_{1:T}^{(L+1)}$,
an approximation of $\varphi_{1:T}^{(L+1)}$,
using the previous approximation~$\hat \varphi_{1:T}^{(L)}$ as input:
\begin{align*}
  \hat \varphi_t^{(L+1)} & \approx \refer G_t \refer M_{t+1}(\hat \varphi_{t+1}^{(L)})
  \\
  & \dots
  \\
  \hat \varphi_T^{(L+1)} & \approx \refer G_T
\end{align*}
The learned auxiliary Feynman-Kac model $(\hat M_{1:T}^{(L+1)}, \hat \eta_{1:T}^{(L+1)})$ follows:
\begin{align*}
  \hat M_{t}^{(L+1)} & = \frac{\hat \varphi_t^{(L+1)}}{\refer M_t(\hat \varphi_t^{(L+1)})} \refer M_t
  ,                                                                                                  \\
  \hat \eta_{t-1}^{(L+1)} & = \refer M_{t}(\hat \varphi_{t}^{(L+1)})
  .
\end{align*}

The simplified \autoref{algo:fwd_th} summarizes how functions $\hat\varphi_t^{(L)}$ are computed recursively.
Since the scheme is forward, the target at each step is available directly from previously computed quantities.
In particular, $\hat \varphi_t^{(L)}$ does not depend on any future approximation $\hat \varphi_{t+h}^{(L)}$,
such that the algorithm is a succession of forward passes over the data.
Furthermore, with fixed depth $L$ of the training,
we can implement an online version of this algorithm that does exactly the same computations,
the construction and some implementation details can be found in \autoref{ann:fwd_online},
including the online version \autoref{algo:fwd_online_th} of \autoref{algo:fwd_th}.
In practice, the approximations depend on a certain learning procedure,
$\Algo^{\mathrm{fwd}}$, which we describe in the next section.

\subsection{Learning the twisting functions from particle samples}

\begin{algorithm}
  \caption{Forward Iterated SMC}
  \label{algo:fwd}
  \KwIn{
    A reference Feynman-Kac model $(\refer \M_{1:T}, \refer G_{1:T})$,
    a number of particles $N$, a number of iterations $L^{\max}$.
  }
  \KwOut{Approximations $\hat \varphi_{1:t}^{(L)}$ for $1\leq L \leq L^{\max}$ of the forward scheme.}
  \BlankLine
  \Set $\hat\varphi_{1:T}^{(0)} \equiv 1$,
  $\hat \M_1^{(0)}\sim\refer \M_1$, $\hat M_{2:T}^{(0)}\equiv \refer M_{2:T}$ and $\hat \eta_{1:T}^{(0)}\equiv 1$.\;
  \For{$L \in 0:L^{\max}-1$}{
    \For{$t \in 1:T$}{
      \tcp{Sample training particles according to previous iterations}
      \Sample $\bar x_{t}^{(L),n}$ and \Compute $\bar w_{t}^{(L),n}$
      according to the $t$-th SMC step (\autoref{algo:SMC})
      with $M_t = \hat M_{t}^{(L)}$ and $\eta_{t-1:t} = \eta_{t-1:t}^{(L)}$
      and ancestors $\ancestor x_{t-1}^{(L+1),1:N}$.\;
      \tcp{Compute the approximations}
      \Set $f(x_{t-1}, x_t)= \refer G_t(x_{t-1}, x_t) \hat\eta_t^{(L)}(x_t)$\;
      \Set $\hat \varphi_t^{(L+1)} = \Algo_t^{\mathrm{fwd}}(f,
      \ancestor x_{t-1}^{(L+1), 1:N}, \bar x_t^{(L), 1:N}, \bar w_t^{(L), 1:N})$.\;
      \Set $\hat\eta_{t-1}^{(L+1)} = \refer M_t(\hat\varphi_t^{(L+1)})$.\;
      \Set $\hat M_t^{(L+1)} = \frac{\hat\varphi_t^{(L+1)}}{\hat\eta_{t-1}^{(L+1)}}\refer M_t$.\;
      \tcp{Sample particles according to trained proposal}
      \Sample and \Compute $x_{t}^{(L+1),n}$, $w_{t}^{(L+1),n}$ and $a_t^{(L+1), n}$
      according to the $t$-th SMC step (\autoref{algo:SMC})
      with $M_t = \hat M_{t}^{(L+1)}$, $\eta_{t-1:t} = \eta_{t-1:t}^{(L)}$
      and ancestor $\ancestor x_{t-1}^{(L+1),n}$.\;
      \Denote $\ancestor x_{t}^{(L+1), n} = x_{t}^{(L+1),a_{t}^{(L+1), n}}$. \;
    }
  }
\end{algorithm}

We use a learning procedure $\Algo^{\mathrm{fwd}}$, in \autoref{algo:fwd}, similar to Controlled SMC
(see \autoref{ann:csmc}),
but the sampling part is more involved.
Particles are sampled as we construct the approximation of the scheme:
for each time $t$ we sample and weight training particles ($\bar x_{t}^{(L),1:N}$, $\bar w_t^{(L),1:N}$)
according to the current auxiliary Feynman-Kac model $(\hat \M_{1:T}^{(L)}, \hat \eta_{1:T}^{(L)})$.
After computing an approximation $\hat\varphi_t^{(L+1)}$ from these particles,
we sample new particles and ancestors ($x_t^{(L+1),1:N}$, $a_t^{(L+1),1:N}$, $w_t^{(L+1),1:N}$)
with the trained proposal $\hat M_t^{(L+1)}$ and the same fixed auxiliary weights $\hat \eta_{1:T}^{(L)}$.

At each step $t$ of iteration $L+1$ of \autoref{algo:fwd}, ancestor particles $\ancestor x_{t-1}^{1:N}$ with indices $a_{t-1}^{1:N}$
and particles $\bar x_t^{1:N}$
weighted by $\bar w_t^n$ from the previous iteration ($L$) are used to build approximations as:
\begin{equation*}
  \hat \varphi_t^{(L+1)}
  \\= \Algo_t^{\mathrm{fwd}}(
    \refer G_t\hat\eta_t^{(L)},
    \ancestor x_{t-1}^{(L+1), 1:N}, \bar x_t^{(L), 1:N}, \bar w_t^{(L), 1:N}
  )
\end{equation*}
where, for a function $f_t$ that will be approximated in the function class $\Psi_t$,
particles $x_{t-1}^{1:N}$ and $x_t^{1:N}$, and weights $w_t^{1:N}$,
$\Algo_t^{\mathrm{fwd}}$ is defined as follows:
\begin{multline*}
  \Algo_t^{\mathrm{fwd}}(f_t, x_{t-1}^{1:N}, x_t^{1:N}, w_t^{1:N})
  \\*
  = \arg \min_{\hat f \in \Psi_t}\sum_{n=1}^N w_t^n\bigg(
    \log\hat f(x_{t-1}^n, x_t^n)
  *- \log f_t(x_{t-1}^n, x_t^n)\bigg)^2
  .
\end{multline*}

When the function class $\Psi_t$ consists of single log-quadratic functions
(i.e.,\ $\Psi_t = \{ x \mapsto \exp(- \frac 1 2 x' A x - x' b - \frac 1 2 c) \mid A \in \mathbb S_d, b\in \R^d, c\in \R\}$),
the minimization above amounts to a simple linear regression problem
(with a semi-definite constraint that is addressed by doing a projection
of the unconstrained solution)
and is relatively easy to compute.

The optimization problem described by $\Algo_t^{\mathrm{fwd}}$ is ill-defined
when the importance weights $w_t^{1:N}$ are too degenerate.
A measure of degeneracy is the effective sample size (ESS):
\begin{equation*}
  \mathrm{ESS}(w^{1:N})
  = \frac{\left(\sum_n w^n\right)^2}
  {\sum_n \left(w^n\right)^2}
  .
\end{equation*}
It represents an effective number of different particles in the sample
and ranges from $1$ (highly degenerate) to $N$ (optimal).
Whenever the ESS falls below $N_0 = 2d$,
where $d$ is the dimension of the space of functions $F$,
the following tempered update is used instead:
\begin{equation*}
  \Algo^{\mathrm{fwd}^{\alpha}}(\dots) = \arg \min_{\hat f \in \Psi_t}
  \sum_{n=1}^N e^{\alpha \log( w_t^n)}
  \bigg(
    \\*
  \log\hat f(x_{t-1}^n, x_t^n)- \log f_t(x_{t-1}^{A_t^n}, x_t^n)\bigg)^2
\end{equation*}
where $\alpha\in(0, 1)$ is a tempering weight. The exponent $\alpha$
is calibrated to ensure that the
ESS of $(e^{\alpha \log(w_t^n)})^{1:N}$ is approximately $N_0$.

The \autoref{algo:fwd} can be used with either
$\Algo_{1:T}^{\mathrm{fwd}}$ or $\Algo_{1:T}^{\mathrm{fwd}^\alpha}$,
but will definitely be less robust without the tempering weight.

The computational cost of \autoref{algo:fwd} is comparable to that of the backward \autoref{algo:cSMC},
although the Forward Iterated SMC algorithm samples roughly twice as many particles as controlled SMC.
If the number of iterations is fixed in advance,
or multiple iterations are performed at once,
an online version of \autoref{algo:fwd} can be used, reusing particles
for both sampling and training, achieving a computational cost comparable to controlled SMC.
These computational cost considerations and online versions of the algorithm
are detailed in \autoref{ann:fwd_fast}.

\section{Numerical experiments}
In this section, we compare our forward-only training algorithm with alternative methods.

We first consider a simple one-dimensional nonlinear model, illustrating the difficulties encountered by backward algorithms.
Then, we study a real data application of the multivariate stochastic volatility
used in \citet{guarniero_iterated_2017}.
Primarily, the methods used for comparison are controlled SMC \citep{heng_controlled_2020}
as a backward training algorithm
and the bootstrap particle filter as a baseline reference.

Our primary performance criterion is an estimate of the variance of the log-normalizing-constant estimator,
i.e.,\ $\mathrm{Var}[\log \hat Z_T]$
(marginal log-likelihood in the bootstrap case).
This quantity serves as a proxy for the relative variance $\var[][\hat Z_T]/\expect[][\hat Z_T]^2$,
which motivates the methods but is harder to estimate directly.
Moreover, it is a commonly used criterion
for tuning the number of particles in particle Metropolis-Hastings algorithms
\citep{Doucet2015, sherlock_optimal_2016}.

We also consider the variance of the importance weights as a measure of degeneracy.
High weight variance indicates high variability in the estimators.
More precisely, we consider the normalized quantity:
\begin{equation*}
  \frac 1 T \sum_{t=1}^T \frac{\frac 1 N \sum_{n=1}^N (w_t^n)^2 - (\frac 1
  N\sum_{n=1}^N w_t^n)^2}{(\frac 1 N\sum_{n=1}^N w_t^n)^2}.
\end{equation*}

\subsection{Nonlinear observation model}
In this section, we are interested in a simple state space model with nonlinear observations,
with latent space $\spaceX = \R$ and observation space $\spaceY = \R$,
and the following autoregressive transition model:
\begin{gather*}
  x_1 = \frac{\sigma_x}{\sqrt{1 - \alpha^2}}\varepsilon_{x,1}
  \\
  x_t = \alpha x_{t-1} + \sigma_x \varepsilon_{x,t}
  ,
\end{gather*}
with $\varepsilon_{x,t} \sim \mathcal N(0, 1)$.
The nonlinear observation model is defined as follows:
\begin{gather*}
  y_t = f(x_t) + \sigma_y \varepsilon_{y, t}
\end{gather*}
with $\varepsilon_{y,t} \sim \mathcal N(0, 1)$
and $f$ is a non-decreasing function.

The choice of $f$ controls the level of difficulty for SMC inference.
For example, if $f$ is a linear function, we get a linear Gaussian model
for which the backward path can be easily and exactly computed.

We set $f(x) = \exp(x) + x/10$ as an example.
With this function, the model is more informative with larger values of $y$.
The intuition is that for small $y$'s, the optimal path will be more difficult to learn and to infer from.

The model is motivated by the observation that most studies in the literature consider
non-linear transitions with linear (or quasi-linear) observations.
Often, the class of functions approximates well the optimal twisting functions in this case.
Non-linear observations, which are very common in applications, introduce additional difficulty due to potentially ill-behaved emission densities.

We try different scenarios with parameter values of
$\alpha\in[0.9, \allowbreak 0.95, \allowbreak 0.98, \allowbreak 0.99, \allowbreak 0.995]$,
$11$ values for $\sigma_x^2$ ranging from $0.05$ to $0.15$
and $6$ values for $\sigma_y^2$ ranging from $0.005$ to $0.055$.
For each scenario, we generated $10$ simulated datasets (over $T=100$ time points).
We ran each algorithm on each dataset and compared their performance.
More precisely, we compare the first iterations of \autoref{algo:fwd} based on the forward scheme
and controlled SMC (\autoref{algo:cSMC}) based on the backward scheme.
We ran each algorithm $64$ times to evaluate the variances of the estimator with $N=1024$.
For each iteration and each run of the algorithm,
we get an SMC output $(x_{1:T}^{1:N}, a_{1:T}^{1:N}, w_{1:T}^{1:N})$ and an associated estimator $\hat Z_T$.

\begin{figure}
  \centering
  \includegraphics[width=\linewidth]{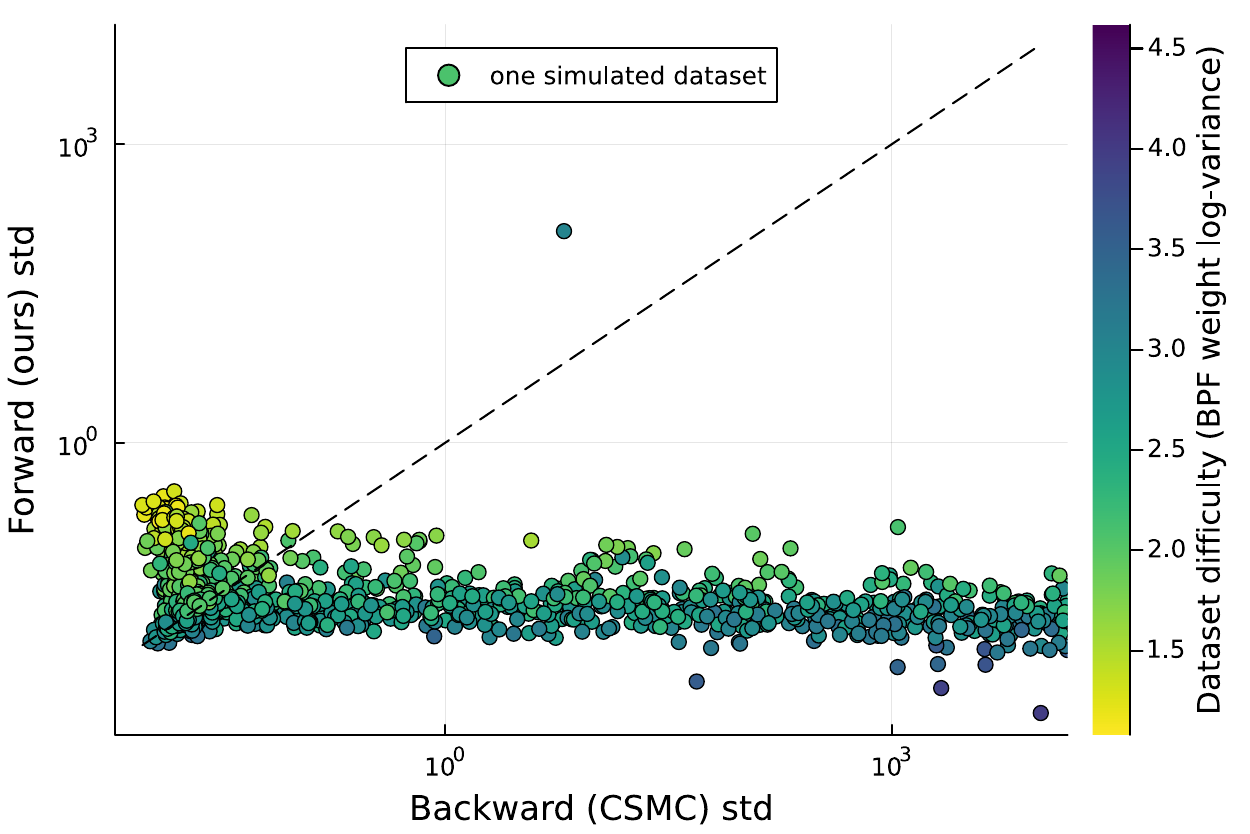}
  \caption{
    Comparison of the standard deviation of $\hat Z_t$ over $64$ runs
    after $L=4$ training iterations
    between a forward algorithm (ours) and controlled SMC \citep{heng_controlled_2020}.
    Better performance is associated with lower standard deviation.
    Each point represents a single simulated dataset
    colored by the mean (over $64$ iterations)
    of the sum of the relative empirical weight variances
    of the bootstrap particle filter (BPF).
    This variance is expected to be lower for an easier dataset.
  }
  \label{fig:nlobs_std_comp}
\end{figure}

\begin{figure}
  \centering
  \includegraphics[width=\linewidth]{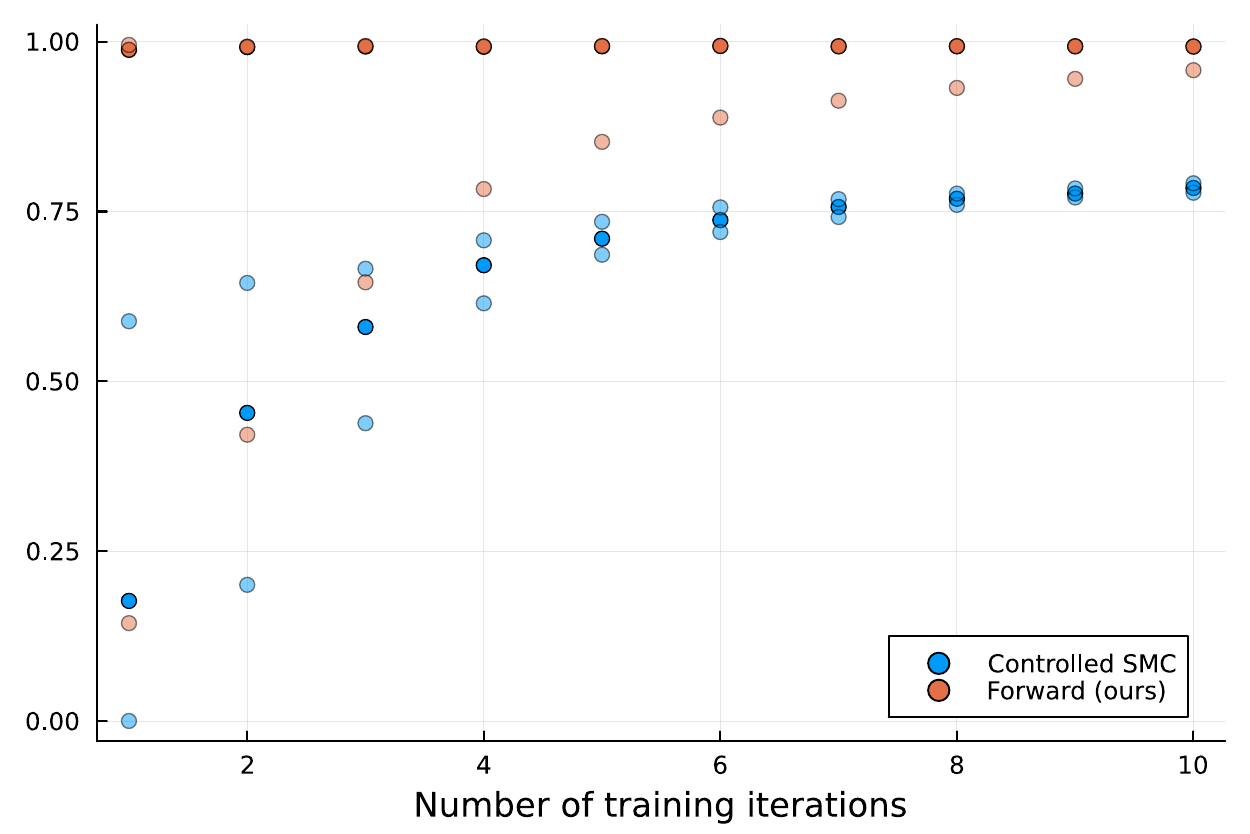}
  \caption{
    Proportion of simulated datasets on which the standard deviation of $\hat Z_t$ over $64$ runs
    is lower than with the bootstrap particle filter (BPF) after each iteration of the training algorithms (from $1$ to $10$).
    Denoting $\text{sd}^{\text{BPF}}$ the standard deviation of $\hat Z_t$ for the BPF,
    lighter dots show $P(\text{sd}^{\text{alg}} \le 0.1\text{sd}^{\text{BPF}})$,
    the proportion of simulated datasets for which the standard deviation of $\text{alg}$ is $10$ time lower
    than the BPF standard deviation (below the main dots),
    and $P(\text{sd}^{\text{alg}} \le 10\text{sd}^{\text{BPF}})$ (above the main dots).
  }
  \label{fig:nlobs_iter}
\end{figure}

We measure "dataset difficulty" by the variance of the bootstrap particle filter weights.
Large weight variance indicates that the dataset is challenging to process.
The \autoref{fig:nlobs_std_comp} shows that the backward algorithm tends to fail
(i.e.,\ to have a high standard deviation)
more often than our forward algorithm,
especially in harder data and parameter sets.
In easy cases, i.e., when the bootstrap particle filter does well,
our forward algorithm does not improve on the backward algorithm.
This is expected since low weight variances suggest less informative data
and slower convergence of the forward scheme toward the optimal scheme.
The \autoref{fig:nlobs_iter} shows that for a low number of iterations, the backward scheme will often return something
that is worse than the bootstrap particle filter (for 80\% of the datasets at iteration $L=1$).
The backward scheme is $10$ times worse than the bootstrap particle filter in about 25\% of the datasets across all iterations.
In most of these cases, the actual trained Feynman-Kac model will end up being either
highly degenerate or will provoke ill-defined inversions in the training.
With the forward scheme, we also observe these issues, but for only $20$ datasets (over $3300$).
This showcases the forward robustness to the dataset
and robustness to the number of iterations: with a lower number of iterations,
we do not make things worse than the bootstrap.

\subsection{Multivariate Stochastic Volatility model}
\label{section:MSV}

In this section, we study the application of our forward-only approach
to a multivariate stochastic volatility model, on a real dataset.
We use the same model and data as \citet{guarniero_iterated_2017}.
We observe data $y_{1:T}$ consisting of monthly log returns on the exchange rate of $d$ currencies.
We infer the stochastic volatility $x_{1:T}$ of dimension $d$ of this log-return as follows:
\begin{equation*}
  y_{t,i}\mid x_t \sim \mathcal{N}(0, \exp(x_{t,i})).
\end{equation*}
The latent space follows an autoregressive transition model:
\begin{align*}
  x_1 & \sim \mathcal{N}(m, \Sigma_\infty )
  \\ x_t \mid x_{t-1} &\sim \mathcal{N}((I - \mathrm{diag} (\alpha))m + \mathrm{diag} (\alpha) x_{t-1}, \Sigma)
\end{align*}
with parameters $m \in \R^d$, $\alpha \in \R^{d-1}$ and $\Sigma \in \mathcal \R^{d\times d}$ positive definite,
and with $\Sigma_\infty$ the covariance matrix such that we sample from the invariant measure at time $1$.

To reduce the dimensionality of the parameter space, \citet{guarniero_iterated_2017} constrain $\Sigma$ to have variances $\sigma^2 \in \R^d$
and zero correlations except for the sub- and super-diagonals, denoted $\rho
\in \R^{d-1}$. We proceed similarly.

We use data from the Federal Reserve\footnote{\url{https://www.federalreserve.gov/releases/h10/hist/}}
from March 2000 to August 2008 ($T=102$).
Like in \citet{guarniero_iterated_2017},
we set a uniform improper prior on $m$, $[0,1]$ uniform priors on each coordinate of $\alpha$,
inverse gamma priors with mean $0.2$ and variance $1$ on each coordinate of $\sigma^2$,
and a $[-1,1]$ triangular prior on each coordinate of $\rho$.
To infer parameters, we do $1.2 \times 10^5$ iterations of particle marginal Metropolis-Hastings (PMMH) on this model
that uses estimators $\hat Z_t$ obtained from running an SMC algorithm.

\begin{figure}
  \centering
  \includegraphics[width=\linewidth]{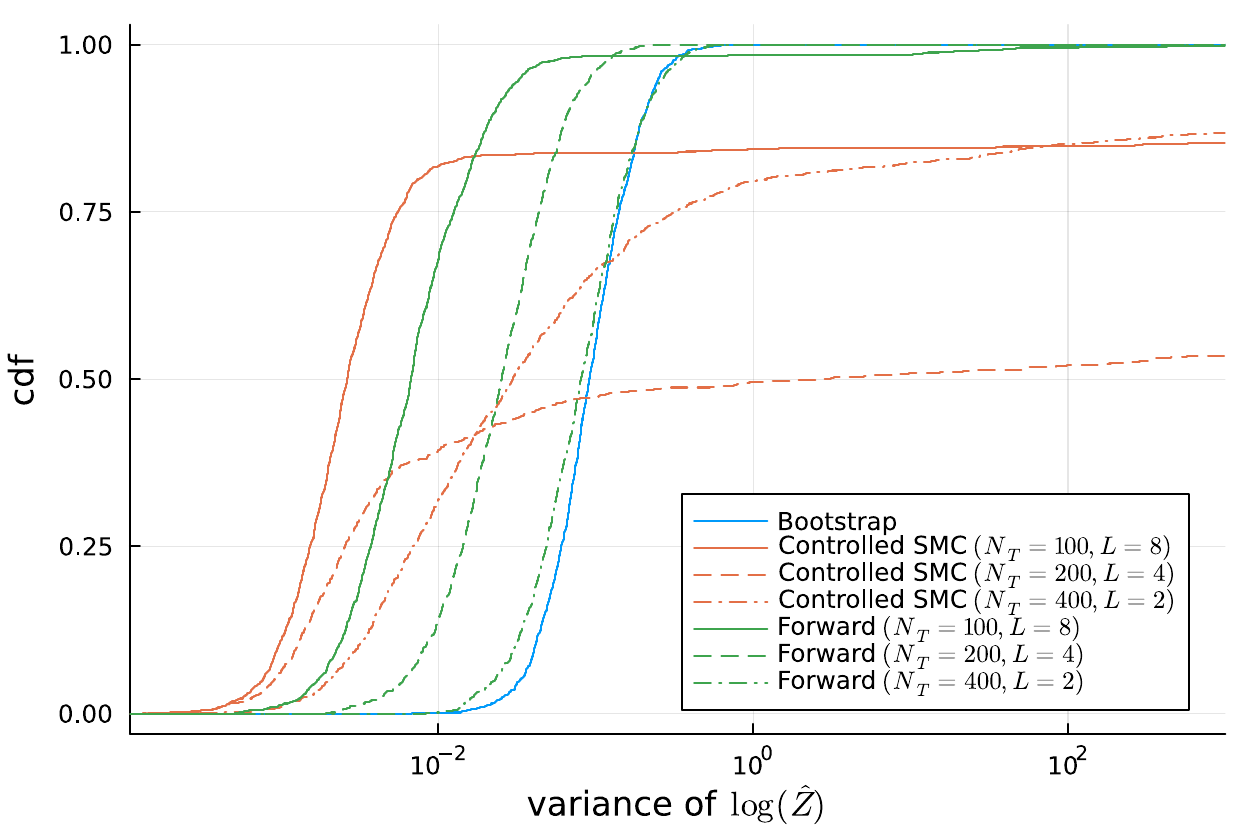}
  \caption{
    Empirical cumulative distribution of the empirical variances of $\log \hat Z_T$
    for $d=8$,
    measured every $100$ steps of the PMMH algorithm over $10$ evaluation.
  }
  \label{fig:msv_xp1_cumul}
\end{figure}

The different methods of estimating $\hat Z_t$ that we compare are:
the bootstrap particle filter with $N$ particles as a baseline and
an SMC algorithm using $\Nsample$ particles with a twisted Feynman-Kac model as input,
trained on the class of log-quadratic twisting functions \eqref{eq:logquad}
with a diagonal quadratic term matrix $A$.
For the training, we use either the controlled SMC algorithm (\autoref{algo:cSMC})
or our forward iterated SMC algorithm (\autoref{algo:fwd}),
both algorithms use $L$ iterations and $\Ntrain$ particles.
We try to match the computational cost of the forward-based SMC algorithm and
the bootstrap particle filter.

We use PMMH with Gaussian random-walk proposals tuned using the forward algorithm.
Beforehand, we transform the model parameters to ensure that proposals respect the parameter constraints.
We initialize the PMMH with the empirical variance of $y_{1:T}$ for $m$,
$0.9$ for $\alpha$, $0.2$ for $\sigma^2$ and $0.25$ for $\rho$.

To assess the variability of the estimates,
we perform $10$ additional SMC runs every $100$ PMMH steps (incurring an extra $10\%$ computational cost) to estimate the empirical variance of $Z_t$.

In the experiment, we studied $d=8$ currencies. We chose $N=4500$ for the bootstrap particle filter,
$\Nsample=600$, and either $(L=8,\Ntrain=100)$, $(L=4,\Ntrain=200)$, or $(L=2, \Ntrain=400)$
such that we use the same number of training particles in each algorithm.
Trace plots indicate that the PMMH chains have reached stationarity,
and the autocorrelation function (ACF) suggests adequate tuning.
In \autoref{fig:msv_xp1_cumul} we observe the empirical CDF
of the estimated variances of $\log \hat Z_T$,
measured every $100$ PMMH steps.
We recall that, when we perform PMMH, we want $\var{}[\log \hat Z_T] \ll 1$.

Our forward iterated SMC algorithm achieves significantly lower variance than the bootstrap particle filter.
Moreover, as observed on simulated data, the forward algorithm proves more robust than the backward algorithm:
while the backward algorithm can potentially reach lower variances, it frequently produces very large variances,
which can cause PMMH chains to become trapped in regions of parameter space where
$\hat Z_t$ estimates are unreliable.
At a fixed training cost, the backward algorithm is also more sensitive to the number of iterations than the forward one.
The behavior of controlled SMC for $(L=2, \Ntrain=200)$ is explained by the chain getting stuck in a high variance region.
A further comparison on the robustness of the forward and backward algorithms on a slightly different dataset is presented in \autoref{ann:msv}.

\appendix

\section{Expression of the variance}
\label{ann:variance}

If we translate equations of \citet{doucet_tutorial_2011} into our notations we get:
\begin{equation}
  \frac{\var{}[\hat Z_T]}{Z_T^2}=
  \frac 1 N \sum_{t=1}^T \bigg(
    \\
  \int \frac{\nu_T(dx_{1:t})}{\nu_{t-1} M_t(dx_{1:t})}\nu_T(dx_{1:t}) - 1\bigg)
  \\
  + O\left(\frac 1 {N^2}\right)
  \label{eq:doucet}
  .
\end{equation}

We define the chi-squared divergence between two probability measures $\nu$ and $\mu$ as:
\begin{equation*}
  \chitwo(\nu \mid \mu)
  = \int \left(\frac{d\mu}{d\nu}\right)^2 d\nu - 1
  = \int \frac{d\mu}{d\nu} d\mu - 1
  .
\end{equation*}
We can express \eqref{eq:doucet} with the chi-squared divergence as follows:
\begin{equation*}
  \frac{\var{}[\hat Z_T]}{Z_T^2}=
  \frac 1 N \sum_{t=1}^T \chitwo(\nu_{t-1}M_t \mid \nu_T)
  + O\left(\frac 1 {N^2}\right)
  .
\end{equation*}

Let us introduce the chi-squared divergence of a non-negative function
$f$ relative to a measure $\mu$ as:
\begin{equation*}
  \chitwo[\mu](f)
  = \chitwo\left(\frac{f}{\mu(f)}\mu \,\middle|\, \mu\right)
  = \int \frac 1 f d\mu \int f d\mu - 1
  .
\end{equation*}
A crucial property of this quantity is that it is null if and only if $f$ is constant,
just like the chi-squared divergence is null when $\nu = \mu$.

Since, for a Feyman-Kac model $(\M_{1:T}, G_{1:T})$,
\begin{align*}
  \nu_{t-1} M_{t}(dx_{1:t})
  & = \frac{1}{Z_{t-1}} G_{1:t-1}(x_{1:t-1}) M_{1:t}(dx_{1:t})
  \\
  \nu_T(dx_{1:t})
  & = \frac{1}{Z_T} G_{1:t}(x_{1:t}) H_{t\to T}(x_t) M_{1:t}(dx_{1:t})
\end{align*}
we have
\begin{equation*}
  \frac{\var{}[\hat Z_T]}{Z_T^2}=
  \frac 1 N \sum_{t=1}^T \chitwo[\nu_T](G_t H_{t\to T})
  + O\left(\frac 1 {N^2}\right)
  .
\end{equation*}

For an auxiliary Feynman-Kac model $(\M_1, M_{2:T}, \eta_{1:T})$, we have:
\begin{equation*}
  G_t H_{t\to T} = \frac{\refer G_{t} \refer H_{t\to T}}{\eta_{t-1}}\frac{d\refer M_t}{dM_t}
\end{equation*}
so we get the following expression:
\begin{equation*}
  \frac{\var{}[\hat Z_T]}{Z_T^2}
  = \frac 1 N \sum_{t=1}^T \chitwo[\nu_T]\left(
    \frac{\refer G_{t} \refer H_{t\to T}}{\eta_{t-1}}\frac{d\refer M_t}{dM_t}
  \right)
  \\
  + O\left(\frac 1 {N^2}\right)
  .
\end{equation*}

\section{Backward algorithm: Controlled SMC}
\label{ann:csmc}

\begin{algorithm}
  \caption{Controlled SMC (2020)}
  \label{algo:cSMC}
  \KwIn{
    A reference Feynman-Kac model $(\refer \M_{1:T}, \refer G_{1:T})$,
    a number of particles $N$, a number of iterations $L^{\max}$.
  }
  \KwOut{Approximations $\hat \varphi_{1:t}^{\star, (L)}$ for $1\leq L \leq L^{\max}$
  of the optimal twisting functions $\varphi_{1:t}^\star$.}
  \BlankLine
  \tcp{Sample initial training particles}
  \Sample ($x_{1:t}^{0,1:N}$, $w_{1:T}^{0,1:N}$, $a_{1:T}^{0,1:N}$)
  from the reference Feynman-Kac model $(\refer \M_{1:T}, \refer G_{1:T})$
  with \autoref{algo:SMC}.\;
  \Denote $\ancestor x_{t}^{(L+1), n} = x_{t}^{(L+1),a_{t}^{(L+1), n}}$. \;
  \For{$L$ \From $1$ \To $L^{\max}$}{
    \tcp{Compute the approximations}
    \Set $\hat\eta_T^{\star, (L)}\equiv 1$.\;
    \For{$t$ \From $T$ \To $1$}{
      \Set $f(x_{t-1}, x_t) = \refer G_t(x_{t-1}, x_t)\eta_t^{\star, (L)}(x_t)$.\;
      \Set $\hat\varphi_t^{\star, (L)} = \Algo^{\mathrm{bck}}(f, \ancestor x_{t-1}^{(L-1), 1:N}, x_{t}^{(L-1), 1:N})$.\;
      \Set $\hat\eta_{t-1}^{\star, (L)} = \refer M_t(\hat\varphi_t^{\star, (L)})$.\;
      \Set $M_t^{\star, (L)} = \frac{\hat\varphi_t^{\star, (L)}}{\hat\eta_{t-1}^{\star, (L)}}\refer M_t$.\;
    }
    \tcp{Sample particles according to the last iteration}
    \Sample ($x_{1:t}^{(L),1:N}$, $w_{1:T}^{(L),1:N}$, $a_{1:T}^{(L),1:N}$)
    from the auxiliary Feynman-Kac model $(\hat \M_{1:T}^{\star, (L)}, \hat \eta_{1:T}^{\star, (L)})$
    with \autoref{algo:SMC}.\;
    \Denote $\ancestor x_{t}^{(L+1), n} = x_{t}^{(i+1),a_{t}^{(l+1), n}}$. \;
  }
\end{algorithm}

In this section, we will explain the Controlled SMC algorithm
developed by \citet{heng_controlled_2020}.

The backward scheme \eqref{eq:backschm}, in terms of twisting functions, yields:
\begin{align*}
  \varphi_T^\star & = \refer G_T
  \\
  & ...
  \\
  \varphi_t^\star & = \refer G_t \refer M_{t+1}(\varphi_{t+1}^\star)
  .
\end{align*}
The algorithm constructs an approximation with twisting functions $\hat\varphi_t^\star$.
The successive approximations follow:
\begin{align*}
  \hat \varphi_T^\star & \approx \refer G_T
  \\
  & ...
  \\
  \hat \varphi_t^\star & \approx \refer G_t \refer M_{t+1}(\hat \varphi_{t+1}^\star)
  .
\end{align*}
The learned auxiliary Feynman-Kac model $(\hat M_{1:T}, \hat \eta_{1:T})$ follows:
\begin{gather*}
  \hat M_t = \frac{\hat \varphi_t}{\refer M_t(\hat \varphi_t)} \refer M_t
  , \quad
  \hat \eta_{t-1} = \refer M_t(\hat\varphi_{t})
  .
\end{gather*}

The backward scheme is not naturally iterative.
The idea of \citet{guarniero_iterated_2017} and \citet{heng_controlled_2020} is that each iteration of the algorithm
samples from an auxiliary Feynman-Kac model $(\hat \M_{1:T}^{\star,(L)}, \hat \eta_{1:T}^{\star, (L)})$
and gives us, applying the backward scheme, the next iteration $(\hat \M_{1:T}^{\star,(L+1)}, \hat \eta_{1:T}^{\star,(L+1)})$.
The idea is that each of the iterations will bring $U_t$
closer to the optimal $U_t^\star$ and the SMC output closer to being directly sampled from the optimal $\nu_T^\star$.
The validity of this convergence was proved under strong assumptions in \citet{heng_controlled_2020}.

Since the backward scheme starts from the end, the corresponding \autoref{algo:cSMC}
builds approximations from outputs of an SMC algorithm $(x_{1:T}^{1:N}, \allowbreak a_{1:T}^{1:N}, \allowbreak w_{1:T}^{1:N})$
obtained with \autoref{algo:SMC} on the auxiliary Feynman-Kac model
$(\hat \M_{1:T}^{\star,(L+1)}, \hat \eta_{1:T}^{\star,(L+1)})$ returned by the previous iteration.
Such outputs provide points used to make the approximations in \autoref{algo:cSMC}.
Thus, if the initial SMC algorithm is degenerate, there is a good chance that the approximation
will not learn anything useful from the target.

Approximations are built as follows:
\begin{equation*}
  \hat\varphi_t^{\star, (L)} = \Algo^{\mathrm{bck}}(
    G_t\hat\eta_t^{\star, (L)},
    x_{t-1}^{(L-1), a_{t-1}^{(L-1),1:N}}, x_{t}^{(L-1), 1:N}
  )
\end{equation*}
where, given a function $f_t$ that is approximated in a set of functions $\Psi_t$, particles $x_{t-1}^{1:N}$ and $x_t^{1:N}$,
$\Algo^{\mathrm{bck}}$ is defined as the minimizer of the following averaged loss over the particles:
\begin{equation*}
  \Algo^{\mathrm{bck}}(f_t, x_{t-1}^{1:N}, x_t^{1:N}) =
  \\ \arg \min_{\hat f \in F}
  \sum_{n=1}^N\bigg(
  \log \hat f(x_{t-1}^n, x_t^n) - \log f_t(x_{t-1}^n, x_t^n)\bigg)^2
  .
\end{equation*}

\section{Online version of the Forward Iterated Algorithm}
\label{ann:fwd_online}

\begin{algorithm}
  \caption{Online SMC Training}
  \label{algo:fwd_online_th}
  \KwIn{
    A reference Feynman-Kac model $(\refer \M_{1:T}, \refer G_{1:T})$,
    a number of particles $N$, a number of iterations $L^{\max}$
    and sets of strictly positive functions $\Psi_{1:T}$.
  }
  \KwOut{Approximations $\hat \varphi_{1:t}^{(L)}$ for $1\leq L \leq L^{\max}$ of the forward scheme.}
  \BlankLine
  \Set $\hat\varphi_{1:T}^{(0)} \equiv 1$,
  $\hat \M_1^{(0)}\sim\refer \M_1$, $\hat M_{2:T}^{(0)}\equiv \refer M_{2:T}$ and $\hat \eta_{1:T}^{(0)}\equiv 1$.\;
  \For{$\tmax \in 1:T$ \KwWith $\tmin = \min(1, \tmax-L^{\max}+1)$}{
    \For{$t$ \From $\tmax$ \To $\tmin$ \KwWith $L=\tmax -t$}{
      \Approximate $\refer G_t \hat\eta_t^{(L)}$ by $\hat \varphi_t^{(L+1)} \in \Psi_t$.\;
      \Set $\hat\eta_{t-1}^{(L+1)} = \refer M_t(\hat\varphi_t^{(L+1)})$.\;
      \Set $\hat M_t^{(L+1)} = \frac{\hat\varphi_t^{(L+1)}}{\hat\eta_{t-1}^{(L+1)}}\refer M_t$.\;
      \tcp{Sample particles according to trained proposal}
      \Sample and \Compute $x_{t}^{(L+1),n}$, $w_{t}^{(L+1),n}$ and $a_t^{(L+1), n}$
      according to the $t$-th SMC step (\autoref{algo:SMC})
      with $M_t = \hat M_{t}^{(L+1)}$, $\eta_{t-1:t} = \eta_{t-1:t}^{(L)}$
      and ancestor $\ancestor x_{t-1}^{(L+1),n}$.\;
      \Denote $\ancestor x_{t}^{(L+1), n} = x_{t}^{(L+1),a_{t}^{(L+1), n}}$. \;
    }
  }
\end{algorithm}

The forward algorithm (\autoref{algo:fwd_th}) can be transformed into \autoref{algo:fwd_online_th}
by changing the order of the operations and iterating over $t$ in the inner loop.
This algorithm is online since at iteration $t$ we can discard all previous $y_{1:t-L^{\max}}$ data.

One important distinction between the forward and the online versions of our method
is that we need to fix the depth $L^{\max}$ of our scheme before we run the online \autoref{algo:fwd_online_th}.
In the forward \autoref{algo:fwd_th}, we can decide to increment $L^{\max}$
at the end of each inner loop.

\section{Computational cost of the Forward Iterated Algorithm}
\label{ann:fwd_fast}

\begin{algorithm}
  \caption{Faster Online Forward Iterated SMC}
  \label{algo:fwd_fast}
  \KwIn{
    A reference Feynman-Kac model $(\M_1, M_{2:t}, G_{1:t})$,
    a number of particles $N$, a number of iterations $L^{\max}$.
  }
  \KwOut{Approximations $\hat \varphi_{1:t}^{(L)}$ for $1\leq L \leq L^{\max}$ of the forward scheme.}
  \BlankLine
  \Set $\hat\varphi_{1:T}^{(0)} \equiv 1$,
  $\hat \M_1^{(0)}\sim\M_1$, $\hat M_{2:T}^{(0)}\equiv M_{2:T}^{(0)}$ and $\hat \eta_{1:T}^{(0)}\equiv 1$.\;
  \For{$\tmax\in 1:T$ \KwWith $\tmin = \max(1, \tmax-L^{\max}+1)$}{
    \tcp{Sample particles}
    \uIf{$\tmin=1$ \KwWith $L=\tmax$}{
      \Sample $x_1^{(L), n} \sim \hat \M_1^{(L)}(dx_t)$.\;
      \Compute $\bar w_1^{(L), n} = \hat G_1^{(L)}(x_t^{(L), n})$
    }\ElseIf{$t=\tmin$ \And $L = L^{\max}$}{
      \Resample $x_{t-1}^{(L), 1:N}$ and get ancestors $a_{t-1}^n$
      according to $w_t^{(L), 1:N}$.\;
      \Sample $x_t^{(L),n}\sim \hat M_t^{(L)}(\bar x_{t-1}^{(L),a_{t-1}^n}, dx_t)$.\;
      \Compute $\bar w_t^{(L),n} = \hat G_t^{(L)}(x_{t-1}^{(L),a_{t-1}^n}, x_t^{(L),n})$. \;
    }
    \For{$t$ \From $\tmin+1$ \To $\tmax$ \KwWith $L= \tmax -t$}{
      \Resample $x_{t-1}^{(L+1), 1:N}$ to $\bar x_{t-1}^{(L+1), 1:N}$
      according to $\bar w_t^{(L), 1:N}$.\;
      \Sample $x_t^{(L), n}\sim\hat M_t^{(L)}(\bar x_{t-1}^{(L+1), n}, dx_t)$.\;
      \Compute $\bar w_t^{(L),n} = \hat G_t^{(L)}(x_{t-1}^{(L+1), n}, x_t^{(L),n})$. \;
    }
    \Set $w_t^{(0), \tmax} = \bar w_t^{(0), \tmax}$ \;
    \tcp{Compute the approximations}
    \For{$t$ \From $\tmax$ \To $\tmin$ \KwWith $L= \tmax -t$}{
      \Set $f(x_{t-1}, x_t)= G_t(x_{t-1}, x_t) \hat\eta_t^{(L)}(x_t)$\;
      \Set $\hat \varphi_t^{(L+1)} = \Algo^{\mathrm{fwd}}(f,
      \bar x_{t-1}^{(L+1), 1:N}, x_t^{(L), 1:N}, \bar w_t^{(L), 1:N})$.\;
      \Set $\hat\eta_{t-1}^{(L+1)} = M_t(\hat\varphi_t^{(L+1)})$.\;
      \Set $\hat M_t^{(L+1)} = \frac{\hat\varphi_t^{(L+1)}}{\hat\eta_{t-1}^{(L+1)}}M_t$.\;
      \Set $\hat G_t^{(L+1)}
      = G_t \frac{dM_t}{dM_t^{(L+1)}}\frac{\eta_t^{(L)}}{\eta_{t-1}^{(L+1)}}$.\;
      \Compute $w_{t-1}^{(L+1)} = \bar w_{t-1}^{(L+1)} \times \frac{\eta_{t-1}^{(L+1)}}{\eta_t^{(L)}}(x_t^{(L),n})$.\;
    }
  }
\end{algorithm}

The computational cost can be divided between two major operations:
sampling particles and solving $\Algo^{\mathrm{fwd}}$ and $\Algo^{\mathrm{bck}}$.
Both the forward \autoref{algo:fwd} and the backward \autoref{algo:cSMC}
use the same number of minimizations per iteration ($T$ minimizations),
but these algorithms do not sample the same number of particles.
At each iteration, the forward \autoref{algo:fwd} uses $N\times T$ sampling of particles for the training
and $N\times T$ sampling of ancestors according to the trained proposal,
whereas the backward \autoref{algo:cSMC} uses the same particles
as training particles and as ancestors in the SMC.

If we denote by $C_{\mathrm{Sample}}$ and $C_{\mathrm{Solve}}$
the cost per particle of these two main operations,
the forward cost $C_{\text{fwd}}$ and the backward cost $C_{\text{bck}}$ are:
\begin{align*}
  \frac{C_{\text{fwd}}}{NT}
  & = 2 C_{\text{Sample}} + C_{\text{Solve}}
  \\
  \frac{C_{\text{bck}}}{NT}
  & = C_{\text{Sample}} + C_{\text{Solve}}
  .
\end{align*}
This assumes that the cost of minimization is linear in $N$ (or with a negligible constant part)
and the same for $\Algo^{\text{fwd}}$ and $\Algo^{\text{bck}}$,
which is the case in our application.

With log-quadratic twisting functions, the sampling cost is linear in the dimension
and the minimization cost is either $O(d^4)$ with a full quadratic term $A$
or $O(d^2)$ with a diagonal quadratic term $A$.
Then, with problems of higher dimension,
the cost per iteration of the forward \autoref{algo:fwd}
is nearly identical to the cost per iteration of the backward \autoref{algo:cSMC}.
Without this log-quadratic setting, the cost per iteration is higher
since it is more involved than the computation of an inverse.
Therefore, the difference in computation time per iteration
should also be small between the algorithms.

One way to reduce the computation time of \autoref{algo:fwd}
is by doing things online.
If one wants to run many iterations of the Forward Iterated Algorithm,
one could recycle the sampled particles to use them for training.
It is done in \autoref{algo:fwd_fast}, which requires half the number of sampling particles
than the initial \autoref{algo:fwd} (exactly the same amount as the backward algorithm).

\section{Details on the robustness of the MSV experiment}\label{ann:msv}

\begin{figure}
  \centering
  \includegraphics[width=\linewidth]{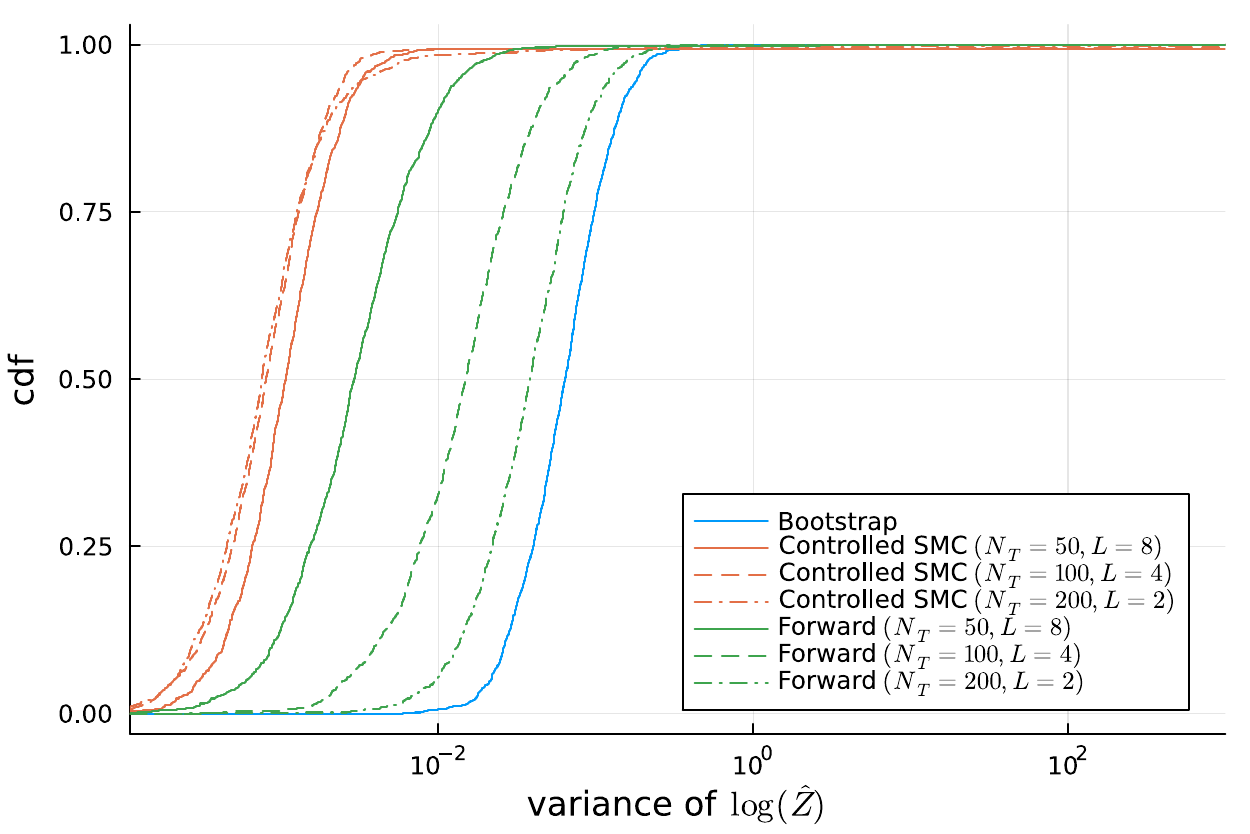}
  \caption{
    Empirical cumulative distribution of the empirical variances of $\log \hat Z_T$
    $d=7$ (bottom),
    measured every $100$ steps of the PMMH algorithm over $10$ evaluation.
  }
  \label{fig:msv_xp2_cumul}
\end{figure}

In this section, we keep the model and settings of the multivariate stochastic volatility model
defined in \autoref{section:MSV}.

We run the same experiment, but we remove one currency (INR -- the Indian rupee),
which is especially difficult for the model.
We end up with $d=7$, we choose $N=3000$ for the bootstrap particle filter,
$\Nsample=800$, and either $(L=8,\Ntrain=50)$, $(L=4,\Ntrain=100)$, or $(L=2, \Ntrain=200)$ for the iterated algorithms.
In \autoref{fig:msv_xp2_cumul}, we observe that all versions of the controlled SMC algorithm
achieve similar variances, better than the forward variances.
By contrast with \autoref{fig:msv_xp1_cumul},
this shows that posterior sampling for this model is not necessarily challenging,
but that real datasets can be more difficult than others.
Controlled SMC, as shown in the simulated experiment (\autoref{fig:nlobs_std_comp}),
does well with easier datasets but struggles with more difficult ones.
Conversely, the forward scheme is slower by nature (especially with easier datasets),
but can work with more difficult datasets, and here gives similar performances in both cases.
This is especially important with real data applications,
since practitioners are not always able to choose their datasets,
and our algorithm extends the sets of models that can be tackled with iterated SMC methods.


\begin{thebibliography}{}

  \bibitem[Andrieu et~al., 2010]{Andrieu2010}
  Andrieu, C., Doucet, A., and Holenstein, R. (2010).
  \newblock Particle markov chain monte carlo methods.
  \newblock {\em Journal of the Royal Statistical Society Series B: Statistical
  Methodology}, 72(3):269–342.
  \newblock \url{https://doi.org/10.1111/j.1467-9868.2009.00736.x}.

  \bibitem[Capp{\'e} et~al., 2005]{cappe_inference_2005}
  Capp{\'e}, O., Moulines, E., and Ryd{\'e}n, T. (2005).
  \newblock {\em Inference in {{Hidden Markov Models}}}.
  \newblock Springer {{Series}} in {{Statistics}}. Springer, New York, NY.
  \newblock \url{https://doi.org/10.1007/0-387-28982-8}.

  \bibitem[Chopin and Papaspiliopoulos, 2020]{chopin_introduction_2020}
  Chopin, N. and Papaspiliopoulos, O. (2020).
  \newblock {\em An Introduction to Sequential {{Monte Carlo}}}.
  \newblock Springer {{Series}} in {{Statistics}}. Springer, Cham.
  \newblock \url{https://doi.org/10.1007/978-3-030-47845-2}.

  \bibitem[Del~Moral, 2004]{del_moral_feynman-kac_2004}
  Del~Moral, P. (2004).
  \newblock {\em Feynman-{{Kac Formulae}}}.
  \newblock Probability and Its {{Applications}}. Springer, New York, NY.
  \newblock \url{https://doi.org/10.1007/978-1-4684-9393-1}.

  \bibitem[Doucet et~al., 2000]{doucet_sequential_2000}
  Doucet, A., Godsill, S., and Andrieu, C. (2000).
  \newblock On sequential {{Monte Carlo}} sampling methods for {{Bayesian}}
  filtering.
  \newblock {\em Statistics and Computing}, 10(3):197--208.
  \newblock \url{https://doi.org/10.1023/A:1008935410038}.

  \bibitem[Doucet and Johansen, 2011]{doucet_tutorial_2011}
  Doucet, A. and Johansen, A.~M. (2011).
  \newblock A tutorial on particle filtering and smoothing : {{Fifteen}} years
  later.
  \newblock In {\em The {{Oxford Handbook}} of {{Nonlinear Filtering}}}, pages
  656--705. Oxford University Press, Oxford ; N.Y.

  \bibitem[Doucet et~al., 2015]{Doucet2015}
  Doucet, A., Pitt, M.~K., Deligiannidis, G., and Kohn, R. (2015).
  \newblock Efficient implementation of markov chain monte carlo when using an
  unbiased likelihood estimator.
  \newblock {\em Biometrika}, 102(2):295–313.
  \newblock \url{https://doi.org/10.1093/biomet/asu075}.

  \bibitem[Doucet and S{\'e}n{\'e}cal, 2004]{doucet_fixed-lag_2004}
  Doucet, A. and S{\'e}n{\'e}cal, S. (2004).
  \newblock Fixed-lag sequential {{Monte Carlo}}.
  \newblock In {\em 2004 12th {{European Signal Processing Conference}}}, pages
  861--864, Vienna, Austria.

  \bibitem[Durbin and Koopman, 2012]{durbin_time_2012}
  Durbin, J. and Koopman, S.~J. (2012).
  \newblock {\em Time {{Series Analysis}} by {{State Space Methods}}}.
  \newblock Oxford University Press, Oxford.
  \newblock \url{https://doi.org/10.1093/acprof:oso/9780199641178.001.0001}.

  \bibitem[Gordon et~al., 1993]{gordon_novel_1993}
  Gordon, N., Salmond, D., and Smith, A. (1993).
  \newblock Novel approach to nonlinear/non-{{Gaussian Bayesian}} state
  estimation.
  \newblock {\em IEE Proceedings F (Radar and Signal Processing)},
  140(2):107--113.
  \newblock \url{https://doi.org/10.1049/ip-f-2.1993.0015}.

  \bibitem[Guarniero et~al., 2017]{guarniero_iterated_2017}
  Guarniero, P., Johansen, A.~M., and Lee, A. (2017).
  \newblock The {{Iterated Auxiliary Particle Filter}}.
  \newblock {\em Journal of the American Statistical Association},
  112(520):1636--1647.
  \newblock \url{https://doi.org/10.1080/01621459.2016.1222291}.

  \bibitem[Heng et~al., 2020]{heng_controlled_2020}
  Heng, J., Bishop, A.~N., Deligiannidis, G., and Doucet, A. (2020).
  \newblock Controlled sequential {{Monte Carlo}}.
  \newblock {\em The Annals of Statistics}, 48(5):2904--2929.
  \newblock \url{https://doi.org/10.1214/19-AOS1914}.

  \bibitem[Kalman, 1960]{kalman_new_1960}
  Kalman, R.~E. (1960).
  \newblock A {{New Approach}} to {{Linear Filtering}} and {{Prediction
  Problems}}.
  \newblock {\em Journal of Basic Engineering}, 82(1):35--45.
  \newblock \url{https://doi.org/10.1115/1.3662552}.

  \bibitem[Olsson et~al., 2008]{olsson_sequential_2008}
  Olsson, J., Capp{\'e}, O., Douc, R., and Moulines, {\'E}. (2008).
  \newblock Sequential {{Monte Carlo}} smoothing with application to parameter
  estimation in nonlinear state space models.
  \newblock {\em Bernoulli}, 14(1):155--179.
  \newblock \url{https://doi.org/10.3150/07-BEJ6150}.

  \bibitem[Pitt and Shephard, 1999]{pitt_filtering_1999}
  Pitt, M.~K. and Shephard, N. (1999).
  \newblock Filtering via {{Simulation}}: {{Auxiliary Particle Filters}}.
  \newblock {\em Journal of the American Statistical Association},
  94(446):590--599.
  \newblock \url{https://doi.org/10.1080/01621459.1999.10474153}.

  \bibitem[Sherlock, 2016]{sherlock_optimal_2016}
  Sherlock, C. (2016).
  \newblock Optimal {{Scaling}} for the {{Pseudo-Marginal Random Walk
  Metropolis}}: {{Insensitivity}} to the {{Noise Generating Mechanism} }.
  \newblock {\em Methodology and Computing in Applied Probability},
  18(3):869--884.
  \newblock \url{https://doi.org/10.1007/s11009-015-9471-6}.

  \bibitem[Xue et~al., 2025]{xue2025onlinerollingcontrolledsequential}
  Xue, L., Finke, A., and Johansen, A.~M. (2025).
  \newblock Online rolling controlled sequential monte carlo.
  \newblock Preprint at \url{http://arxiv.org/abs/2508.00696}.

\end{thebibliography}
\end{document}